\documentclass[apj]{emulateapj}

\usepackage{color}
\usepackage[colorlinks=true,
            linkcolor=black,
            urlcolor=black,
            citecolor=blue, breaklinks=true]{hyperref}
\usepackage{epsfig} 
\usepackage{epstopdf}
\usepackage{graphicx}
\usepackage{amsmath}

\shorttitle{Cluster-scale gradient in ISM metallicity}
\shortauthors{Gupta et al.}

\begin{document}

\hypersetup{linkcolor=blue}

\title{Radial Distribution Of ISM Gas-phase Metallicity In CLASH Clusters at $z\sim 0.35$:  A New Outlook On Environmental Impact On Galaxy Evolution}

\author{Anshu Gupta\altaffilmark{1}, Tiantian Yuan\altaffilmark{1},  Kim-Vy H. Tran\altaffilmark{2}, Davide Martizzi\altaffilmark{3}, Philip Taylor\altaffilmark{1},  and  Lisa J. Kewley\altaffilmark{1}}

\email{anshu.gupta@anu.edu.au}

\altaffiltext{1}{Research School of Astronomy and Astrophysics, Australian National University, Canberra, ACT 2611, Australia}
\altaffiltext{2}{George P. and Cynthia W. Mitchell Institute for Fundamental Physics and Astronomy, Department of Physics \& Astronomy, Texas A\&M University, College Station, TX 77843}
\altaffiltext{3}{Department of Astronomy, University of California, Berkeley, CA 94720-3411, USA}

\begin{abstract}

We present the first observation of cluster-scale radial metallicity gradients from star-forming galaxies. We use the DEIMOS spectrograph on the Keck II telescope to observe two CLASH clusters at $z\sim0.35$: MACS1115+0129 and RXJ1532+3021.  Based on our measured interstellar medium (ISM) properties of star-forming galaxies out to a radius of $2.5\ {\rm Mpc}$ from the cluster centre,  we find that the galaxy metallicity decreases as a function of projected cluster-centric distance ($-0.15\pm0.08\  {\rm dex/Mpc}$) in MACS1115+01.  On the mass-metallicity relation (MZR),  star-forming galaxies in MACS1115+01 are offset to higher metallicity ($\sim0.2\ {\rm dex}$) than the local SDSS galaxies at a fixed mass range.  In contrast, the MZR  of  RXJ1532+30 is consistent with the local comparison sample.  RXJ1532+30 exhibits a bimodal radial metallicity distribution, with one branch showing a similar negative gradient as MACS1115+01 ($-0.14\pm0.05\ {\rm dex/Mpc}$) and the other branch showing a positive radial gradient. The positive gradient branch in RXJ1532+30 is likely caused by either interloper galaxies or an in-plane merger, indicating that cluster-scale abundance gradients probe cluster substructures and thus the  dynamical state of a cluster.  Most strikingly, we discover that neither the radial metallicity gradient nor the offset from the MZR is driven by the stellar mass. We compare our observations with  Rhapsody-G cosmological hydrodynamical zoom-in simulations of relaxed galaxy clusters and find that the simulated galaxy cluster also exhibits a negative abundance gradient, albeit with a shallower slope ($-0.04\pm0.03\ {\rm dex/Mpc}$).  Our observations suggest that the  negative radial gradient originates from ram-pressure stripping and/or strangulation processes in the cluster environments.

\end{abstract}


\section{Introduction}

Understanding the effect of cluster environment on galaxy formation and evolution is a central topic in extragalactic astronomy.  A galaxy falling into the dense environment of a galaxy cluster  experiences many high-speed encounters with the cluster members and tidal interactions with the dark matter halo of the cluster, leading to the removal of gas and disk destabilization \citep[galaxy harassment,][]{Moore1996}. At the same time,  the hot intra-cluster medium (ICM)  interacts with the interstellar medium (ISM)  of the galaxy and  strips off its  gas reservoir through ram pressure \citep[ram pressure stripping,][]{Gunn1972, Dressler1983, Gavazzi1995}. The galaxy loses a large fraction of the halo gas as it moves through the cluster,  resulting in a gradual decline of star formation as it slowly runs out of fuel \citep[strangulation,][]{Larson1980}. Additionally, if the dynamical friction time is sufficiently short, the galaxy will ultimately reach the cluster center and merge with the central brightest cluster galaxy (BCG) \citep[galactic cannibalism,][]{Ostriker1975, Hausman1978, DeLucia2007}.  All of the above processes are directly associated with the  mass and ICM gas distribution \citep[e.g.,][]{Gunn1972, Dressler1980, DeLucia2007, bialas2015}.  The theoretical dark-matter density profile of a cluster is universal \citep[e.g.,][]{Navarro1997, Huss1999}, however, the exact structure of the local mass and ICM gas density depend on the dynamical state of the cluster \citep{Dressler1988, Serna1996, Adami2005}. 

Gas-phase metallicity provides a fossil record of the formation history of galaxies, modulated by galactic-scale gas flows.  Metallicity can trace the impact of mass and environment on galaxy evolution.  The mass-metallicity relation  in the local universe has minimal  dependence on environment as shown by most observations and simulations  of local field and cluster galaxies \citep{Mouhcine2007, Cooper2008, Ellison2009, Dave2011, Scudder2012, T.M.Hughes2012, Pasquali2012}. Galaxies in clusters tend to have slightly higher metallicity ($\sim$0.04 dex) when compared to a control sample of field galaxies \citep{Ellison2009}. However, \cite{Ellison2009} find that the slight enhancement in metallicity is mainly driven by the presence of close  companions, not simply the cluster membership. The metallicity of star-forming satellite galaxies at $z\sim0$ shows a weak positive correlation with environmental over-density; however, such a correlation is not seen in central galaxies \citep{peng2014}. In high density environments such as inner regions of the galaxy clusters at $z\sim0$, the gas phase metallicity is higher by $0.08-0.20$ dex than counterpart field galaxies \citep{Mouhcine2007}. \cite{Darvish2015} found that galaxies in the filamentary structure at $z\sim0.5$ have $\sim0.15$ dex higher metallicity than their counterpart field galaxies. 

 Studies at higher redshift ($z\sim2$) are inconclusive about the  environmental dependence of metallicity \citep[e.g.,][]{Kacprzak2015, Tran2015,Valentino2015, Shimakawa2014}.   Most studies on the environmental dependence of mass-metallicity relation compare galaxies residing in  low versus high-density environment \citep{Mouhcine2007, Cooper2008, Scudder2012}, but the local environment density in a cluster depends on the cluster-centric distance. Therefore, to assess the effect of the cluster environment on chemical evolution, it is essential to consider the cluster-centric distance.

Galaxy properties such as color, morphology and luminosity show a strong dependence on the density (or cluster-centric distance) in galaxy clusters.  The morphology-density relation shows that the elliptical galaxy fraction is higher in the cluster core than the cluster  outskirts \citep[e.g.,][]{Dressler1980, Houghton2015}.   Similarly, the environmental dependence of the luminosity function shows that the most luminous (or most massive) galaxies exist in the densest parts of the universe, i.e., the cores of galaxy groups and clusters \citep{Davis1988,Zandivarez2011}. In addition, the red galaxy fraction decreases with cluster-centric distance \citep{Martinez2008, Ribeiro2013}, implying an increasing star formation rate with cluster-centric distance. Simulations of galaxy clusters can reproduce the color distribution of cluster members by assuming a decreasing star formation rate after galaxies enter the cluster \citep{Balogh2000}. A recent study by \cite{Maier2016} finds a higher fraction of metal rich galaxies in the   accreted region of the galaxy cluster than galaxies in the infalling region. However, there is no comprehensive study on the radial dependence of cluster member metallicities.

The radial metallicity distribution  can constrain the mass assembly of galaxy clusters and the effect of cluster environment on the evolution of  individual  galaxies within the cluster. Both simulations and X-ray observations find that the X-ray plasma in galaxy clusters have higher metallicities  in the cluster core than at larger cluster radius \citep{DeGrandi2001, Hlavacek-Larrondo2013}, representing the excess average star formation in the cluster core. The observed negative radial gradients in ICM metallicity are often used as a constraint in galaxy cluster simulations \citep[e.g.,][]{Kapferer2007, cora2008, Rasia2015, Martizzi2015}. Recent simulations by \cite{Martizzi2015} show that the ICM metallicities can be explained by a simple regulator model, driven by metal enrichment from stellar nucleosynthesis  and gas accretion.  \cite{Martizzi2015, Rasia2015} also find a mild correlation between stellar metallicity and cluster-centric distance, with X-ray luminous (cool-core) clusters exhibiting steeper metallicity gradients than non-cool core clusters.    These results imply that the chemical evolution of individual cluster galaxies is intimately linked with the local ICM properties of the cluster.  Observations of the ISM gas-phase metallicity in galaxies as a function of the ICM density (or cluster-centric distance)  provide new insights into the interplay between  cluster galaxies and the cluster environment. 

 We present the first observations of the radial distribution of global ISM metallicity of star-forming galaxies in galaxy clusters. This work is based on optical spectra obtained with the DEIMOS spectrograph on Keck II telescope. The large field of view (FOV) of the DEIMOS spectrograph allows  the observation of cluster members up to 2 times the virial radius ($\sim 2 \times r_{vir}$) as compared to the small central region probed by X-ray observations ($\sim 200\ {\rm kpc} \sim 0.2\times r_{vir}$). We investigate the global metallicity of star-forming galaxies as a function of radial distance in two X-ray luminous clusters, MACS1115+0129 and RXJ1532+3021 at $z \sim 0.35$, and compare our observations with the Rhapsody-G simulations from \cite{Martizzi2015}. 

We organize this paper as follows. In Section \ref{sec:obs}, we describe the cluster sample selection, observations, data reduction and the derived quantities including the metallicity diagnostics. The main results of this paper are presented in Section \ref{sec:results}.  In Section \ref{sec:discussion}, we discuss the possible mechanisms responsible for the origin of a cluster scale metallicity gradient. Throughout this work, we assume a standard $\Lambda$CDM cosmology with $h = 0.696$,  $\Omega_{\Lambda} = 0.714$ and $\Omega_M = 0.286$ \citep{Hinshaw2013}.

\section{ Observations and data reduction}\label{sec:obs}

\subsection{Cluster Sample Selection}\label{sec:sample}

The data used in this paper are part of our Gravitationally Lensed-galaxies Observed With-Adaptive Optics (GLOW-AO) survey.  For the GLOW-AO survey the galaxy clusters were selected from the   Cluster Lensing And Supernova survey with Hubble \citep[CLASH,][]{Postman2012} clusters. Our GLOW-AO survey was designed to detect lensed galaxies in the redshift range of $0.8<z<1.6$ that can be observed with adaptive optics.  The CLASH galaxy clusters were originally selected from a sample of X-ray bright clusters based on their gravitational lensing strength. The  GLOW-AO survey is optimized for lensed galaxies to be followed up with AO observations using the OSIRIS instrument at the Keck I telescope.  We, therefore select clusters that have at least one bright star ($R_{\rm AB} < 18.0$) within a radius of $60''$ from the central BCG.  In order to perform a comprehensive analysis of the effect of cluster environment on the abundance gradient, we choose galaxy clusters with at least 15 cluster members. This restricted our cluster sample to 2 out of the 8 observed galaxy clusters, MACS1115+0129 and RXJ1532+3021. 

\begin{deluxetable*}{lrrrrrrrr}
\tablecolumns{9}
\tablewidth{0pc}
\tablecaption{X-ray and optically derived properties of galaxy cluster sample \label{tb:clus_info}}
\tablehead{
\colhead{Cluster}  &
 \colhead{$z$\tablenotemark{a}} &
  \colhead{ kT \tablenotemark{b}} &
   \colhead{ $L_{\rm Bol}$\tablenotemark{b} } & 
   \colhead{$r_{\rm vir}$\tablenotemark{c}} &
   \colhead{$M_{\rm vir}$\tablenotemark{c}} &
   \colhead{$c_{\rm vir}$\tablenotemark{c}} &
   \colhead{$\sigma_{cl}$ \tablenotemark{d}} &
   \colhead{$M_{\rm vel}$\tablenotemark{d}} \\
\colhead{}  &
 \colhead{} &
  \colhead{  (keV)} &
   \colhead{ ($10^{44} {\rm\ erg\ s^{-1}}$)} & 
   \colhead{$({\rm Mpc/}h)$} &
   \colhead{ $ (10^{15} M_{\odot}/h)$} &
   \colhead{} &
   \colhead{$({\rm km/ s})$\tablenotemark{d}} &
   \colhead{ $ (10^{15}  M_{\odot}/h)$} 
}
\startdata 
MACS1115+0129 & 0.352 & $8.0\pm0.4$ & $21.1 \pm 0.4$ &  1.78 & $1.13\pm0.10$ & $2.9 \pm 0.9$& 960 $\pm$ 147 & 1.91 $\pm$ 0.61 \\
RXJ1532+3021 & 0.362 & $5.5\pm0.4$ & $20.5\pm0.9$ &  1.47 & $0.64\pm 0.09$ & $3.8 \pm1.7$ & 1487 $\pm$ 213 & 3.78 $\pm$ 0.73\\
\enddata
\tablenotetext{a}{Redshift derived from SDSS spectra of central BCG.}
\tablenotetext{b}{kT is the temperature and  L$_{Bol}$ is the total luminosity derived from the X-ray data \citep{Postman2012}. }
\tablenotetext{c}{$r_{\rm vir}$ is the virial radius, $ M_{\rm vir}$ is the total halo mass at the virial radius, and  $c_{\rm vir}$ is concentration parameter at the virial radius of the cluster  derived by combining the strong and weak lensing observations \citep{Merten2014}.}
\tablenotetext{d}{$\sigma_{cl}$ is the radial velocity dispersion from our observations and $M_{\rm vel}$ is the halo mass estimated from the velocity dispersion.}

\end{deluxetable*}

MACS1115+01 \citep{Ebeling1998} and RXJ1532+30 \citep{Ebeling2010} are  massive clusters with  virial masses of  $\sim 10^{14} M_{\odot}$ \citep{Postman2012}. Table $\sim$\ref{tb:clus_info} lists  the derived physical and observational  parameters for the selected clusters. X-ray morphology maps for both clusters exhibit  minimal disturbances, highlighting the dynamically relaxed nature of the clusters \citep{Allen2007}. However, radio mini-halos were found around the central $\sim$200 kpc of RXJ1532+30, suggesting a possible dynamical disturbance in the central region of this cluster \citep{Kale2013}. No radio disturbance is found in MACS1115+01.  The non-uniform large scale temperature distribution from X-ray observations of RXJ1532+30 suggests a past minor merger with a cooler sub-cluster  \citep{Hlavacek-Larrondo2013}. Spectroscopic data for MACS1115+01 and RXJ1532+30 are very limited. Our optical spectra provide important  data to investigate dynamical structures of the clusters through the velocity and metallicity distributions of cluster members.

\subsection{Local comparison sample}\label{sec:sdss_sample}

For our mass-metallicity study, we derive the local  comparison sample from the Sloan Digital Sky Survey Data Release 7 \citep[SDSS-DR7,][]{York2000, Abazajian2009}. Despite being at a lower redshift,  SDSS galaxies  were used as  the comparison sample as there is a little evolution in the ISM properties  of  galaxies between $z\sim 0.1$ to $z \sim 0.35$ \citep{Kewley2013}.  We use publicly available MPA-JHU catalogs for the stellar masses and emission line flux measurements  of the SDSS DR7 galaxies\footnote{ \label{fn:1} See \url{http://wwwmpa.mpa-garching.mpg.de/SDSS/DR7/Data/stellarmass.html}} \citep{Kauffmann2003a, Brinchmann2004}. 
We use a simple mass cut to match the stellar mass range of the SDSS sample to our DEIMOS cluster data. To ensure $> 20\%$ aperture coverage for each galaxy and minimize the aperture effects \citep{Kewley2005}, we select SDSS galaxies within the redshift range of $0.05<z<0.10$.  We select star-forming galaxies using the standard BPT diagram criteria  described in \cite{Kewley2006}.   We further restrict the SDSS sample by selecting galaxies with signal-to-noise (S/N) ratio $ \ge 5.0$ for the following emission lines,  [OII]$\lambda\lambda 3726,3729$, H$\beta\lambda4861$, [OIII]$\lambda\lambda 4959, 5007$, [NII]$\lambda\lambda 6549,6583$, H$\alpha\lambda 6563$ and [SII]$\lambda\lambda 6717,6731$.  The final sample to compare with our cluster data is 39388 for MACS1115+01 and 47400 for RXJ1532+30.

\subsection{ Mask Design and Spectroscopy}\label{sec:mask}

\begin{deluxetable}{lr}
\tablecolumns{2}
\tablewidth{0pc}
\tablecaption{Priority list of targets \label{tb:prt}}
\tablehead{
\colhead{Objects }  &
 \colhead{Priorities\tablenotemark{a}}
}
\startdata 
Lensed candidates & 1000\\
Photometric $z$ ($0.8<z<2.5$) & 900\\
NGS galaxies  & 800\\
LGS galaxies & 600\\
Galaxies ($R_{\rm AB}<24.0$) & 100\\
\enddata
\tablecomments{Objects represents the category of galaxies on the DEIMOS mask.}
\tablenotetext{a}{Priorities is the number assigned by us during the DEIMOS mask designing.}

\end{deluxetable}%

The mask for MACS1115+01 was designed using publicly available Hubble Space Telescope (HST) images and photometric catalogs.  Because the HST images cover only the central $\sim120''$, we used Canada France Hawaii Telescope (CFHT) color images offered by C. J. Ma and Harald Ebeling (private communication) to select potential cluster members and background lensed galaxies. Lensed galaxy candidates were identified based on their morphology, position around central  BCG and photometry.  The highest priority slits were placed on candidate lensed galaxies. The subsequent priorities of  objects are listed in Table \ref{tb:prt}. Galaxies that can be observed with a natural guide star, i.e. a $R_{\rm AB} < 18.0$ star at a separation of  $<30''$ radius from the  central BCG, are referred to as NGS. Similarly, galaxies referred as LGS can be observed with laser guide star, i.e. they have a star with $R_{\rm AB} < 18.0$ within a $60''$ radius. The remainder of the mask was filled with galaxies with $R_{\rm AB}<24$.  We designed each mask  to have at least  two stars with $R_{\rm AB} < 15$ on each side to ensure proper mask alignment. Our GLOW-AO survey targeted primarily strongly lensed galaxies. The strong lensing region ($\sim$ central 500 kpc) of the cluster mask was filled with lensed galaxy candidates, thus deliberately missing out cluster members in the central part of the cluster.

\subsection{DEIMOS Observations}

All observations were conducted with the DEIMOS spectrograph \citep{Faber2003} at the KECK II telescope on UTC February 24, 2014.  We used  the 600ZD grating at the central wavelength of $7500$\AA\ and the OG550  filter. Our spectra had a wavelength coverage of 5500 \AA\ $-$ 9800 \AA\ with a dispersion of $0.65$ \AA/pixel and spectral resolution of ${\rm R} \sim 2000 $. The slit width for all observations was $1''$ and all slits had a constant position angle the same as the DEIMOS mask. A DA-type white dwarf star G191-B2B \citep{Rauch2013} of $V_{\rm AB}=11.74$ was used for flux calibration. We developed a robust  technique to flux calibrate the DEIMOS observations by deriving the separate sensitivity function of individual CCDs of the spectrograph (Gupta et al.,  submitted).  Absolute flux calibrated spectrum of the star G191-B2B was obtained from the ESO archive\footnote{\label{fn:2} See \url{http://www.eso.org/sci/observing/tools/standards/spectra.html|}}. Table \ref{tb:ob_sm} summarizes the observations for the data used in this paper. The mask for MACS1115+01 was observed with a total integration time of 1 hour split into $3 \times 20$ minutes exposures to eliminate cosmic rays. The mask for RXJ1532+30 had a total integration time of 1 hour 40 minutes ($5\times 20$ minutes exposures). Each mask had $\approx 130$  slits and the average seeing during observations was $\sim 0.8''$.  

\begin{deluxetable*}{lllccc}
\tablecolumns{6}
\tablewidth{0pc}
\tablecaption{Observation summery\label{tb:ob_sm}}
\tablehead{
\colhead{Targets}  &
 \colhead{$\alpha$(J2000)\tablenotemark{a}} &
  \colhead{$\delta$(J2000)\tablenotemark{a}} &
   \colhead{Mask} & 
   \colhead{Airmass} &
\colhead{Exposure time} 
}
\startdata 
MACSJ1115+01 & 11:15:52.1$^a$ & +01:29:53$^a$  & Multi-slit & 1.06 & $3 \times 1200s$\\
RXJ1532+30 & 15:32:53.8$^a$ & +30:20:58$^a$  & Multi-slit & 1.09 & $5 \times 1200s$\\
G191-B2B$^b$ & 05:05:30.6 & +52:49:54 & Long-slit & 1.19 & 45s\\
G191-B2B$^c$ & 05:05:30.6 & +52:49:54 & no mask (CCD1) & 1.19 & 15s\\
G191-B2B$^c$ & 05:05:30.6 & +52:49:54 & no mask (CCD2) & 1.19 & 15s\\
G191-B2B$^c$  & 05:05:30.6 & +52:49:54 & no mask (CCD3) & 1.19 & 15s\\
G191-B2B$^c$  & 05:05:30.6 & +52:49:54 & no mask (CCD4) & 1.19 & 15s\\
\enddata
\tablenotetext{a}{Coordinates of the BCG centroid (J2000).} 
\tablenotetext{b}{Used for telluric correction and longslit mode flux calibration.}
\tablenotetext{c}{Slitless mode flux calibration.}
\tablecomments{Seeing $\sim 0.8''$ }

\end{deluxetable*}%

\subsection{Data reduction}

We reduced the data in  two stages. First, we used the publicly available spec2d pipeline written by Michael Cooper \citep{Cooper2011} to reduce the raw data. The pipeline generates 2D slit-spectra after performing the  bias removal,  flat-fielding, slit-tilt correction, cosmic ray rejection and wavelength calibration. Secondly, we used our IDL-based code (Gupta et al., submitted) to correct for atmospheric absorption, derive instrument sensitivity and flux calibrate the 2D slit spectra. 

For telluric correction, we used the long-slit mode observation of the standard star G191-B2B, which is a DA type white dwarf star with an effective temperature of 60920K  \citep{Rauch2013}.  The atmospheric absorption profile convolved with instrument response curve  was derived from stellar 1D spectrum after removing the intrinsic stellar absorption features and the underlying blackbody profile. All 2D spectra were divided by the derived sensitivity curve to correct for instrument sensitivity and telluric absorption. 

We used the process described in Gupta et. al, (submitted) to flux calibrate 2D slit spectra from DEIMOS.  The 1D-spectrum of the observed objects was extracted by fitting a Gaussian curve to the spatial slit profile and integrating the flux within the full width half maxima (FWHM). For galaxies where Gaussian fitting was not possible, we used FWHM calculated from the point spread function (PSF) of alignment stars on the DEIMOS mask to extract 1D spectrum.

\subsection{Emission line fitting}\label{sec:em_flux}

We visually inspected all 2D slit spectra to estimate an initial redshift by  identifying strong emission lines like [OII] $\lambda\lambda 3726,3729$, H$\beta$, [OIII]$\lambda 4959, 5007$, H$\alpha$, [NII]$\lambda 6583$. To determine the need to correct for stellar absorption before extracting emission line fluxes, we select emission-line dominated galaxies that have sufficient S/N in the continuum to fit stellar absorption spectra (20\% of the emission line sample).  Continuum subtraction was performed with penalised pixel-fitting routine \citep[pPXF,][]{Cappellari2004}  based on the MILES stellar population models \citep{Vazdekis2010}. We find that the difference in line ratios before and after continuum subtraction is largest in log([OIII]/H$\beta$) ($\sim0.05$ dex), whereas for other line ratios the difference is negligible. The difference in N2S2-diagnostic metallicity (See Section \ref{sec:met})  before and after continuum subtraction is less than 0.04 dex, smaller than the statistical error of our observations ($\sim 0.07$ dex). Therefore,  stellar absorption plays a negligible role in the emission-line sample tested above.  A majority ($\sim$ 80\%) of our emission line sample do not have sufficient S/N in the continuum. We do not perform stellar absorption fitting before extracting emission lines for consistent analysis across the entire sample. 

\begin{figure*}
  \centering
  \tiny
\includegraphics[scale=0.50, trim=0.3cm 2.5cm 0.3cm 1.5cm,clip=true]{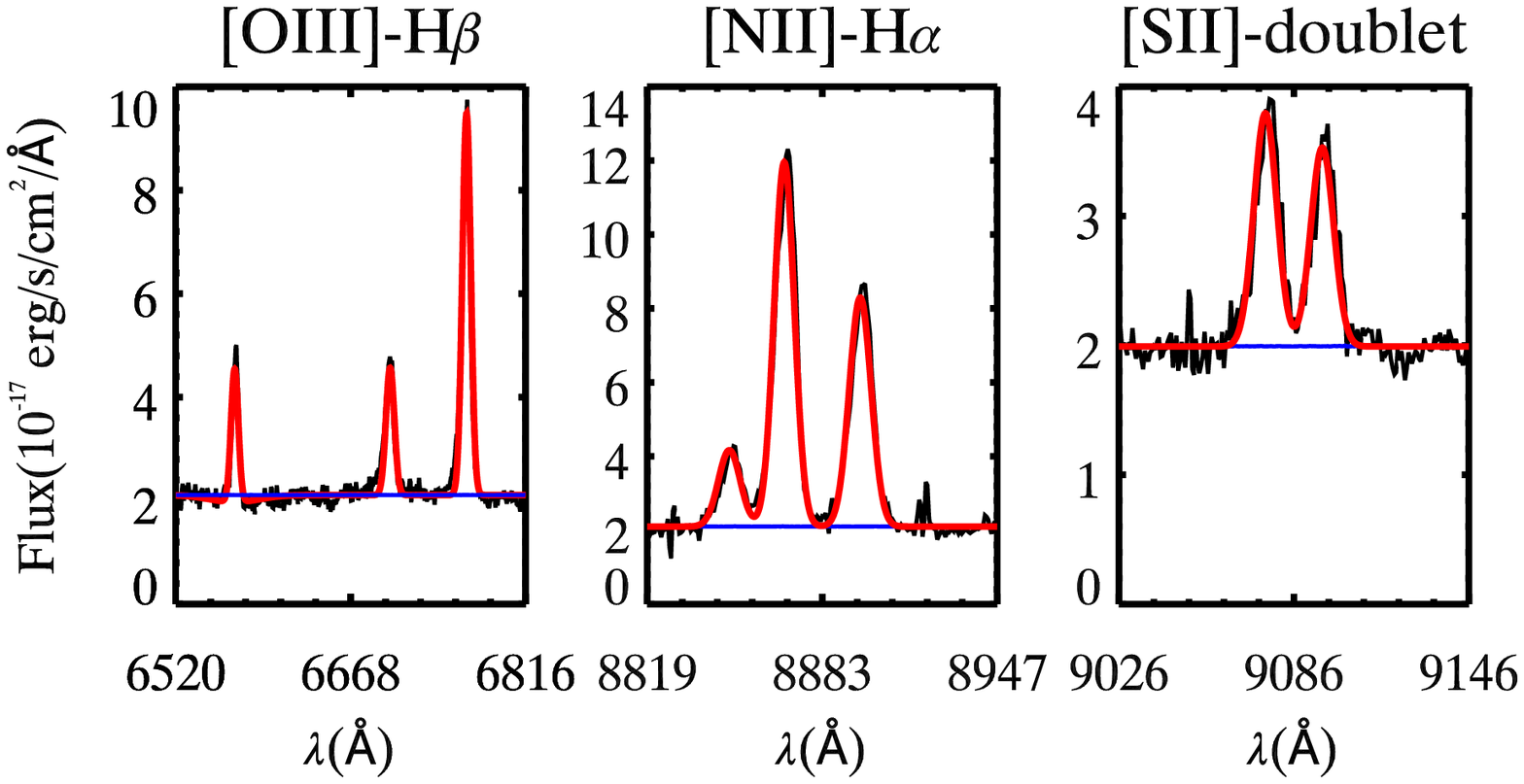}
\includegraphics[scale=0.40, trim=1.0cm 0.0cm 1.0cm 1.0cm,clip=true]{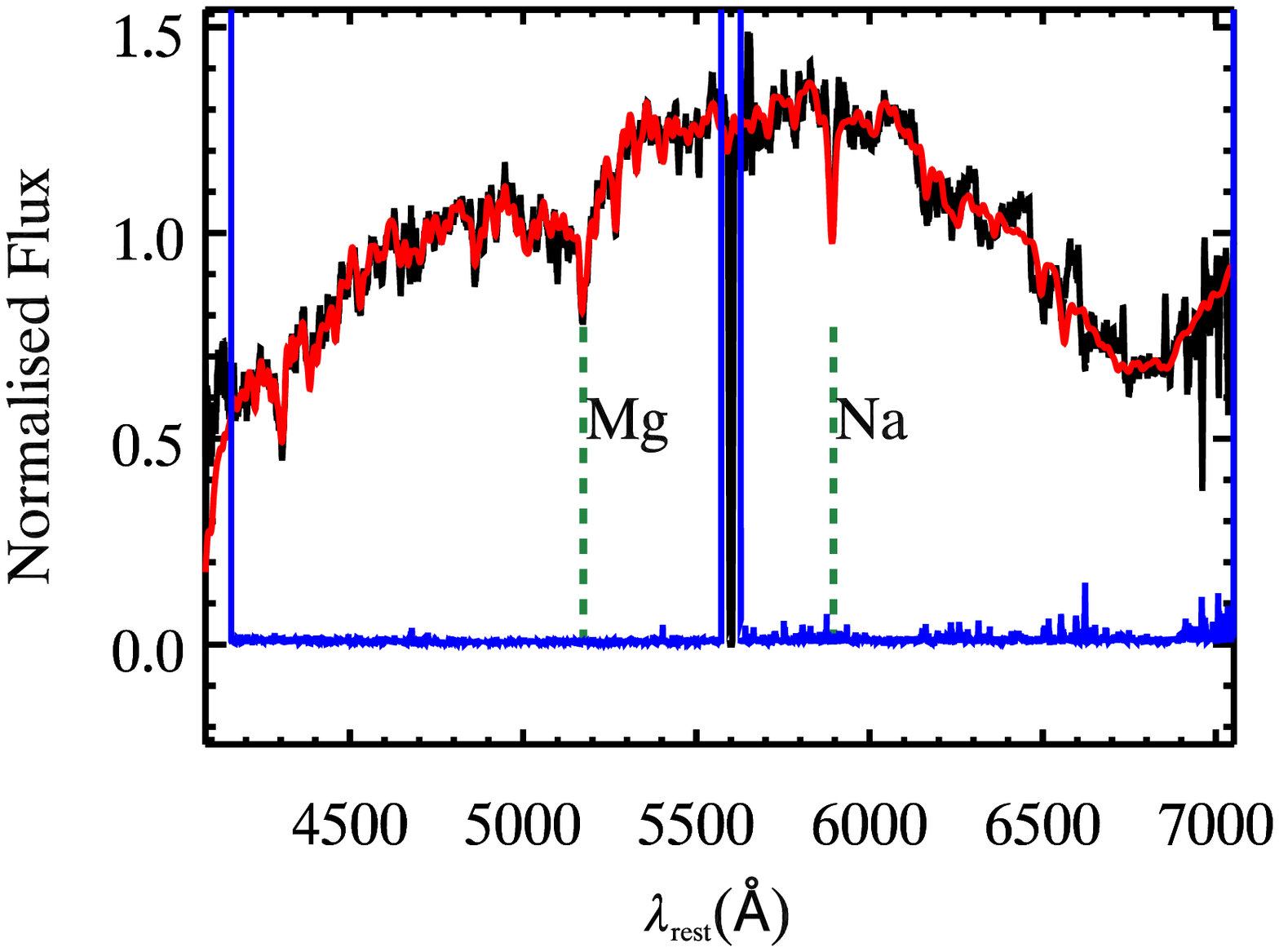}
\caption{DEIMOS spectra of cluster members in MACS1115+01. {\bf Left panel:} An example of emission line galaxy spectrum at $z=0.351$ and line-profile fittings in following order, [OIII]-H$\beta$, [NII]-H$\alpha$ and [SII].  {\bf Right panel:} An example of absorption line galaxy spectrum (black curve) with the pPXF fit (red curve) at $z= 0.347$. Blue curve shows the error spectrum supplied to the pPXF.}
\label{fig:fits}
\end{figure*}

Multiple Gaussian profiles were fit to emission lines simultaneously depending on their separation in wavelength.  For the [OII] doublet $\lambda\lambda 3726,3729$, we fit a double Gaussian with 5 free parameters: redshift, flux peak-1, flux peak-2, line width and continuum level. A similar double Gaussian function was used for the [SII]$ \lambda\lambda 6717,6731$ doublet. We fit a triple Gaussian to the [NII], H$\alpha$ emission lines using 6 free parameters: redshift, flux-H$\alpha$, flux-[NII], line width-H$\alpha$, line width-[NII] and continuum level. The ratio of [NII]$\lambda 6548$ to [NII]$ \lambda6583$ was constrained to its theoretical value of $1/3$ \citep{1989agna.book.....O}.  The [OIII], H$\beta$ emission line group was simultaneously fit using 4 Gaussians with 8 free parameters: redshift, flux-[OIII], width-[OIII], flux-H$\beta$, width-H$\beta$, flux absorption-H$\beta$, width absorption-H$\beta$ and continuum level.  Figure \ref{fig:fits}: left panel presents the example of emission line fits for cluster member in MACS1115+01 at $z=0.351$. We identified 73 redshifts out of 127 slits in the redshift range of $0.05<z<1.13$ for the mask MACS1115+01. For RXJ1532+30, 83 redshifts were identified out of the 139 slits in the redshift range of $0.036<z<0.992$. Ratios of major emission lines, [OIII]/H$\beta$, [NII]/H$\alpha$, [SII]/H$\alpha$ and [NII]/[SII] for the cluster members in MACS1115+01 and RXJ1532+30 are given in Table \ref{tb:data_table_1115} and Table \ref{tb:data_table_1532} respectively.

\subsection{Absorption line galaxies}\label{sec:abs_fit}

Spectra with no apparent emission lines but with a strong continuum were visually inspected for Na$\lambda\ 5872$, Mg$\lambda\ 5174$ and Balmer absorption lines, to estimate an initial redshift. We performed single stellar population fitting using the pPXF to fine-tune  the redshifts of the absorption line galaxies. We used the MILES stellar population models \citep{Vazdekis2010} to best match the spectral resolution of our observations. The skylines and bad pixels  were masked before fitting the continuum models. The right panel in Figure  \ref{fig:fits} shows the example of fit from pPXF for a cluster member in MACS1115+01 at $z=0.347$. Using pPXF, we calculated the redshift of 8 absorption line galaxies for MACS1115+01 (Table \ref{tb:data_table_1115_abs}) and 15 galaxies for RXJ1532+30 (Table \ref{tb:data_table_1532_abs}). 

\subsection{Active galatic nuclei (AGN) removal} \label{sec:agn_removal}

\begin{figure*}
  \centering
  \tiny
\includegraphics[scale=0.46, trim=0.0cm 0.0cm 0cm 0.0cm,clip=true]{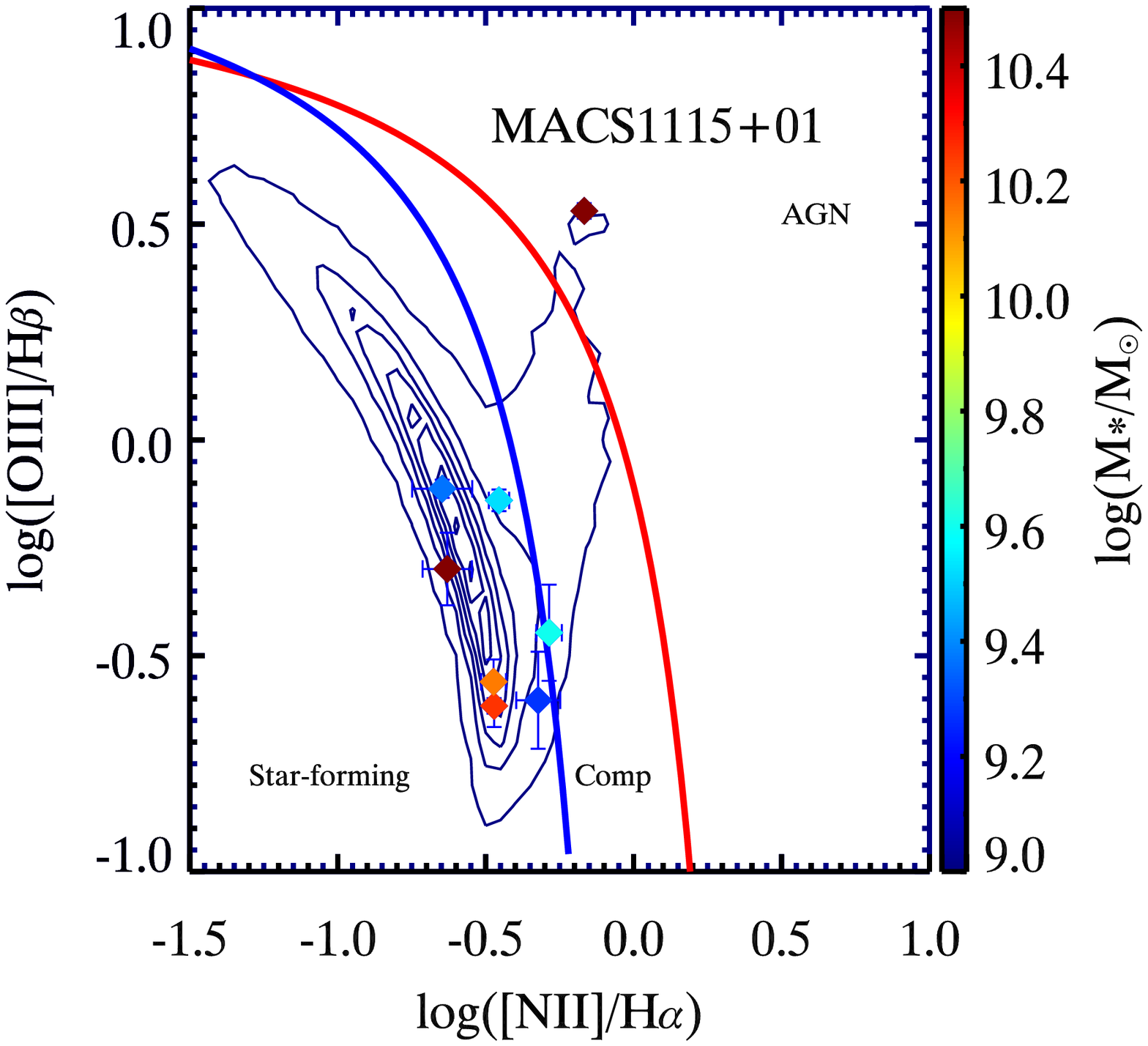}
\includegraphics[scale=0.46, trim=0.0cm 0.0cm 0cm 0.0cm,clip=true]{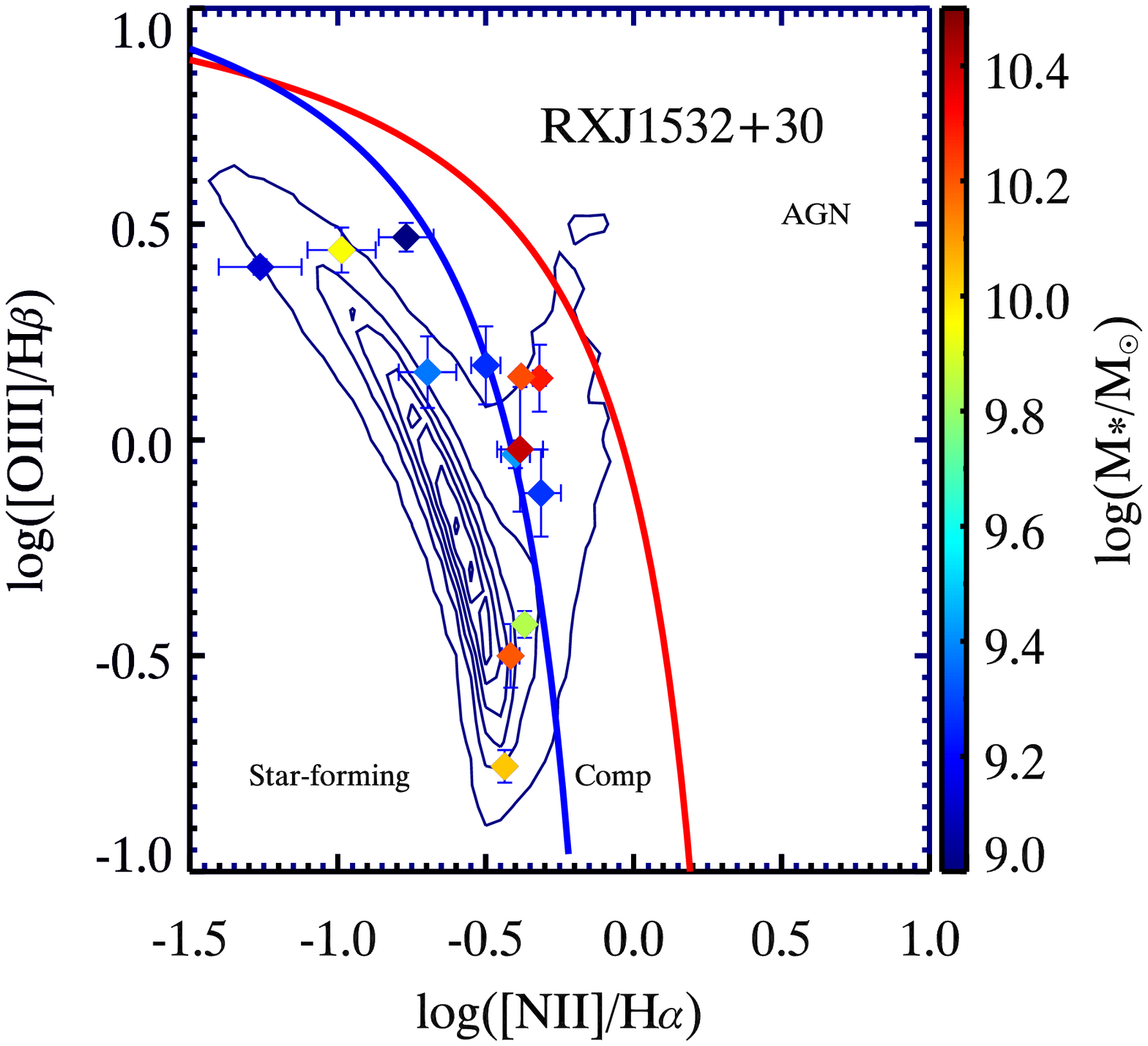}
\caption{Optical  BPT diagnostic diagrams for the observed cluster members for MACS1115+01 (left panel) and RXJ1532+30 (right panel). The data points in each panel are color coded with stellar mass (${\rm \log( M_*/M_{\odot})}$). The red \citep{Kewley2001}  and blue \citep{kauffmann2003} curves divide the galaxies into three subclasses i.e. star forming, composite and AGN.  The blue contours show line ratios for the local comparison sample. AGN galaxies are removed to ensure star forming sample for the optical diagnostics.}
\label{fig:bpt}
\end{figure*}

Abundance calibrations using strong emission lines  are only robust for star-forming galaxies. Therefore, we need to remove AGNs from our cluster member sample.  Galaxies can be classified on the basis of their global gas excitation properties on the Baldwin-Phillips-Terlevich (BPT) diagram \citep{Baldwin1981, Phillips1986, Veilleux1987}. The hard ionizing radiation field from AGN increases the intensity of  the metal ion  forbidden lines (e.g., N$^+$, O$^{++}$)  as compared to the hydrogen recombination lines. 
The presence of large-scale shocks  or aged stellar population in post-starburst galaxies also result in enhanced intensity of emission lines from the metal ions. Therefore, galaxies with the presence of AGN/ evolved hot stars, and/or shocks form a separate group on the BPT diagram than the star-forming galaxies \citep{Kewley2006, Sarzi2010, Rich2011, Yan2012, Alatalo2016}.
\citet[hereafter Ke01]{Kewley2001} developed a theoretical ``maximum starburst line'' on the BPT diagram using a combination of stellar population synthesis, photoionisation and shock models. An ``empirical star-forming line'' was developed by \citet[hereafter Kau03]{kauffmann2003} using the SDSS data. Using Ke01 (red curve) and Kau03 (blue curve), we can separate galaxies into star-forming, composite and AGN groups on the  [O III]/H$\beta$ and [N II]/H$\alpha$ the BPT diagram (Figure \ref{fig:bpt}).   The AGN versus star-forming classification was developed for $z\sim 0$ galaxies. Because the redshift evolution between $z\sim0$ and $z\sim0.35$ is negligible on the BPT diagrams, the $z\sim0$ classification can be used for our cluster sample at $z\sim0.35$. Our results remain unchanged even if we use the new redshift dependent classification by \cite{Kewley2013}. For analysis of metallicity gradients, we use galaxies below the Ke01 line,  thus incorporating both the star-forming and the composite galaxies.

\subsection{Stellar mass}\label{sec:mass}

We derived the stellar mass from  SDSS broad-band photometry ($u, g, r, i, z$) using the {\it Le Phare} package \citep{Arnouts1999, Ilbert2006} with the stellar templates from \cite{Bruzual2003} and Chabrier initial mass funcation \citep[IMF,][]{Chabrier2003}. The stellar templates have seven exponentially decreasing star formation models ($\mathrm{SFR} \propto e^{-t/\tau}$) with $\tau = 0.1, 0.3, 1, 2, 3, 5, 10, 15, 30$ Gyr and three metallicities. To correct for dust attenuation in stellar continuum, the  \cite{Calzetti1999} models were used  with $\mathrm{E(B-V)} = 0\ {\rm to}\ 0.6$. The stellar templates with stellar population ages from 0 to 13 Gyr were used.  Throughout this paper, we have used the median of the generated mass distribution as the stellar mass of our galaxies. The typical error from {\it Le Phare}  in  the mass estimation from the SDSS photometry is 0.2 dex.

\subsection{ISM parameters}\label{sec:met}

We use a new theoretical calibration by \cite{dopita2016} to derive the global metallicities for our selected cluster member galaxies. Direct measurement of the ISM metallicity  requires observations of extremely weak auroral emission lines such as [OIII]$\lambda 4363$, which is difficult for high redshift galaxies. Thus, the metallicity is often determined via theoretical and empirical calibrations based on strong emission lines such as H$\alpha$ and [NII]. The strength of these emission lines depends on the ionization conditions in the galaxies  i.e., ionization parameter and ISM pressure.  The new calibration by \cite{dopita2016} using the $\mathrm{log([NII]/H\alpha)}$ and $\mathrm{log([NII]/[SII])}$ ratios  is independent of ionization conditions and  scales linearly with the oxygen abundance ($12+{\rm log(O/H)}$), a proxy for the total metal content.

We use the \cite{dopita2016} calibration given by:
\begin{equation}
\begin{split}
\mathrm{12 + \log(O/H) = 8.77 + \log([NII]/[SII])} + 
\\ {\rm 0.264\log([NII]/H\alpha)},
\end{split}
\end{equation}
to derive the metallicity for both our cluster galaxies and the local comparison sample. We require a signal to noise (S/N) of 3 for all  necessary emission lines to derive the metallicity.   Using an independent metallicity calibration such as the \citet[][N2]{Pettini2004} does not affect our results. We also determine the ionization properties of the cluster galaxies  using [OIII]/H$\beta$ and [NII]/H$\alpha$. We found no statistically significant relationship between the ionization state of the gas and cluster radius, given our small sample size.

\section{Results}\label{sec:results}
\subsection{Cluster member identification}

\begin{figure*}
  \centering
  \tiny
\includegraphics[scale=0.5, trim=0.0cm 0.0cm 0.0cm 0.0cm,clip=true]{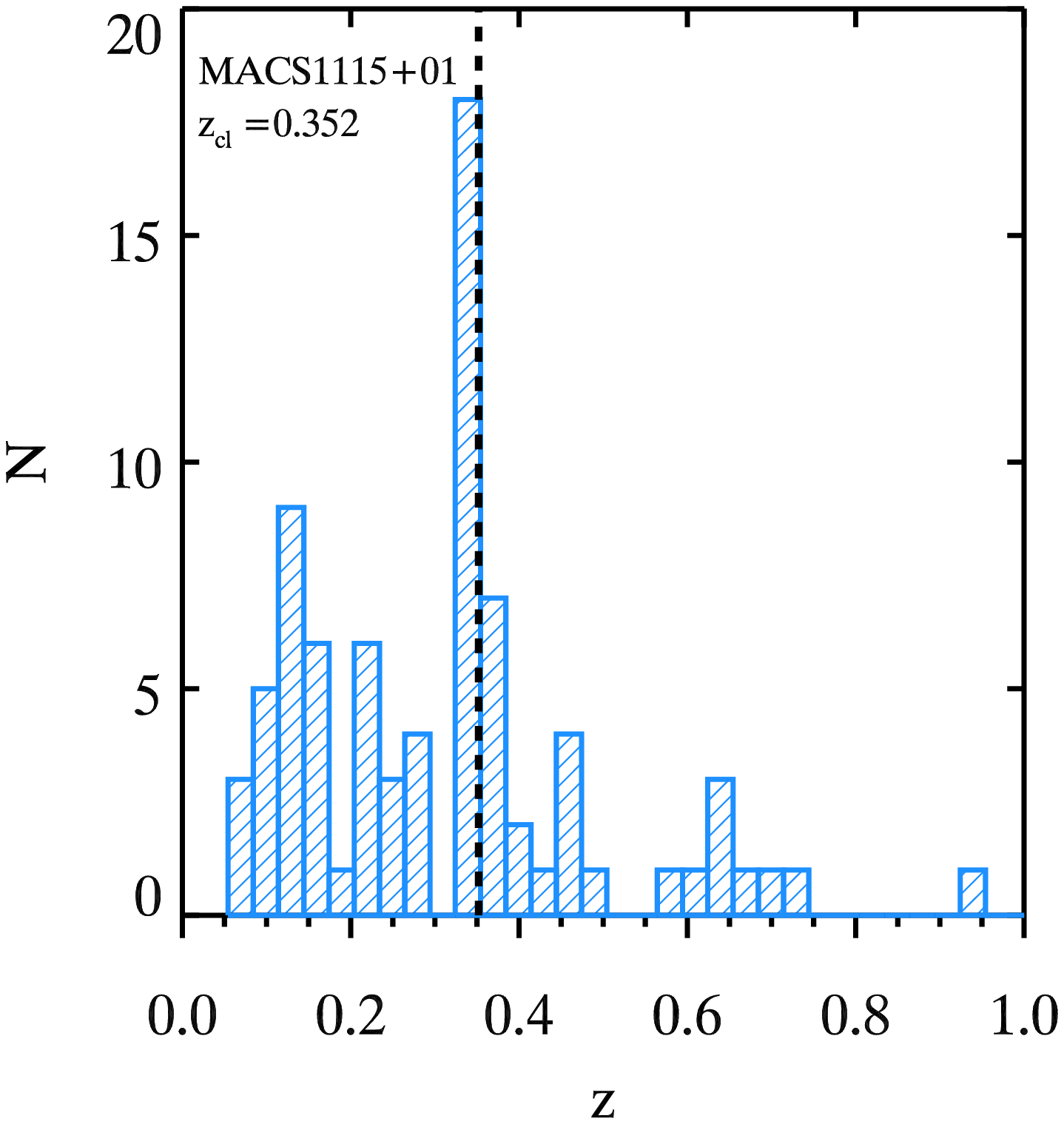}
\includegraphics[scale=0.5, trim=0.0cm 0.0cm 0.0cm 0.0cm,clip=true]{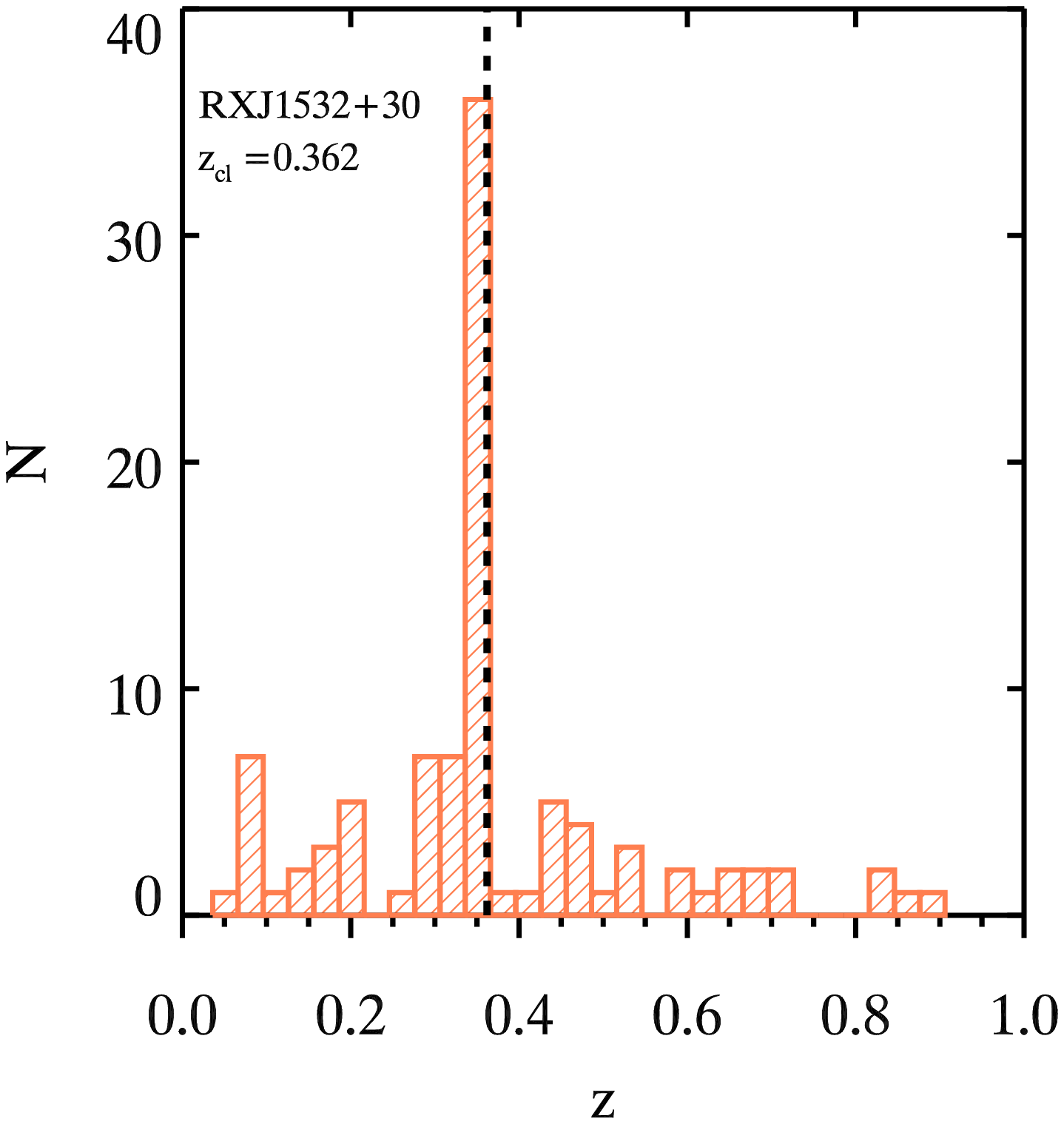}
\caption{Redshift distribution of all objects observed on the DEIMOS mask. {\bf Left panel:} MACS1115+01, and {\bf right panel:}  RXJ1532+30.  The black line in each panel indicates the redshift of the clusters. The peak in histogram at the cluster redshifts clearly indicates detection of cluster members. }
\label{fig:rs_dis}
\end{figure*}

We identified cluster members based on the radial velocity of galaxies around the redshift of the central BCG.  The peak of the redshift distribution for each mask  coincides with the redshift of the central BCG of the respective clusters (Figure \ref{fig:rs_dis}). We used an initial radial velocity cut of  $\pm 10,000 \mathrm{\ km/s}$ and a further $2.5\sigma$ clip on the radial velocity with a basic outlier rejection.

\begin{figure}
  \centering
  \tiny
\includegraphics[scale=0.5, trim=0.0cm 0.0cm 0.0cm 0.0cm,clip=true]{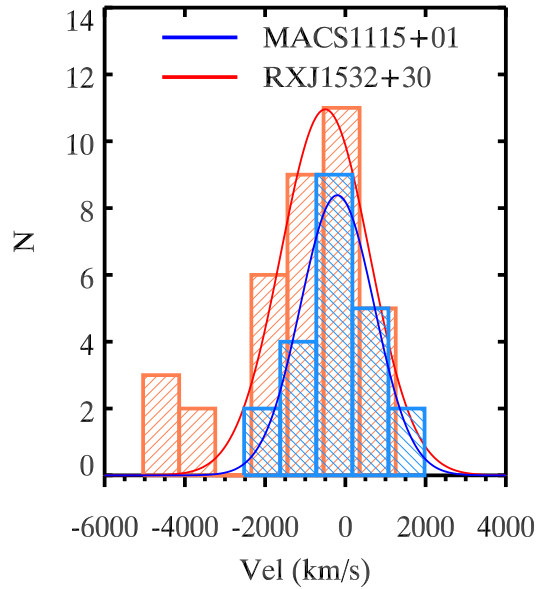}
\caption{The radial velocity distribution of selected cluster members. The cluster members for MACS1115+01 are shown in blue and RXJ1532+30 in red. Solid lines indicates the Gaussian profiles of the velocity distribution clearly showing the different dynamical state of the two cluster.}
\label{fig:rd_dis}
\end{figure}

\begin{figure*}
  \centering
  \tiny
\includegraphics[scale=0.5, trim=0.0cm 0.0cm 0.0cm 0.0cm,clip=true]{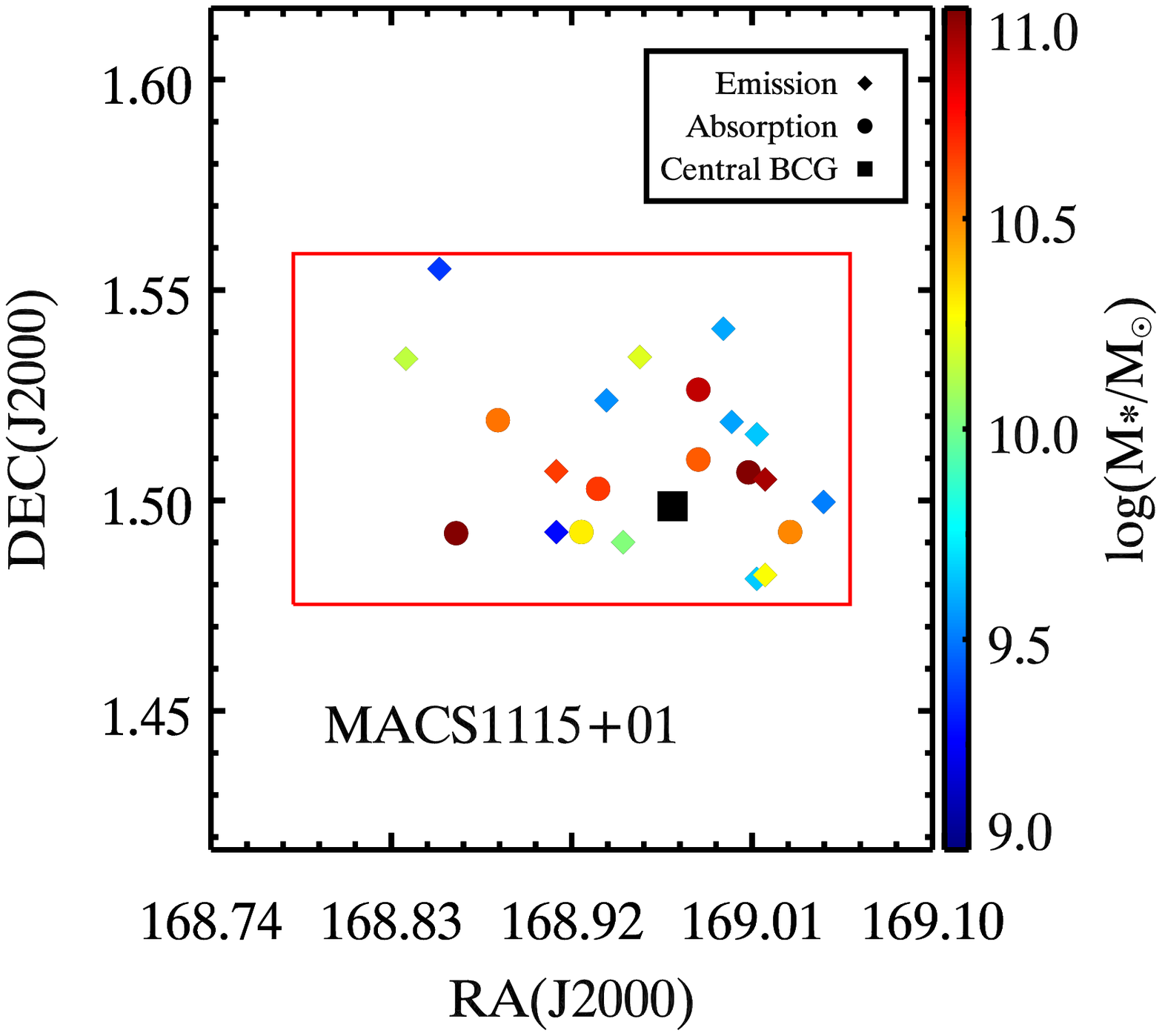}
\includegraphics[scale=0.5, trim=0.0cm 0.0cm 0.0cm 0.0cm,clip=true]{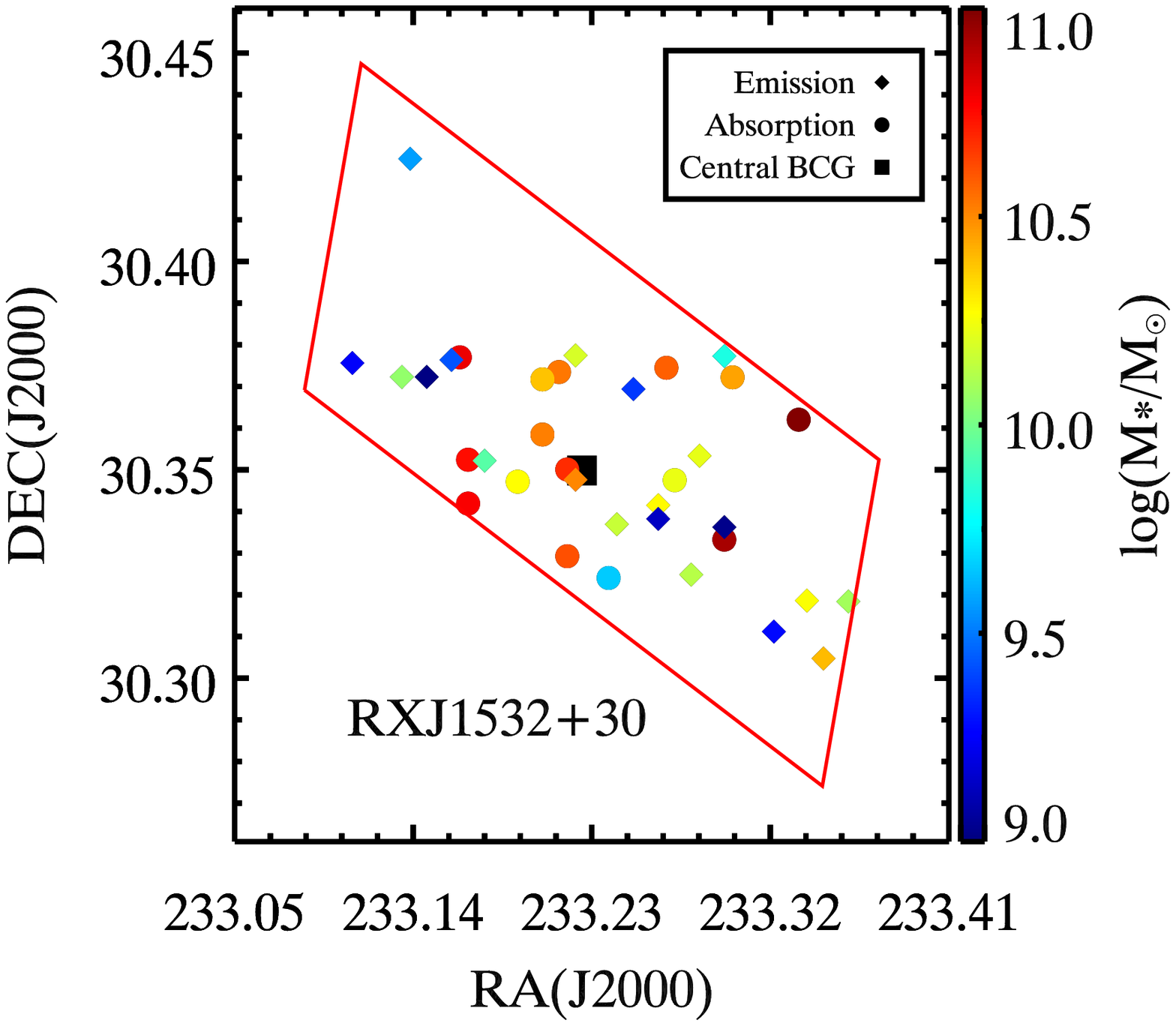}
\caption{Spatial distribution of all cluster members with absorption line galaxies shown as circles and emission line galaxies shown as diamonds. {\bf Left panel:} MACS1115+01, and {\bf right panel:}  RXJ1532+30. The square box in each panel shows the location of the central  BCG for each cluster. The large box in each panel represents the field of view of the DEIMOS mask used in our observations. The data is color coded with stellar mass ($\log({\rm M_*/M_{\odot}})$) through out this paper. This figure does not show evidence of mass segregation for both clusters as the spatial location of cluster member is independent of their stellar mass.}
\label{fig:sp_dis}
\end{figure*}

We identified 22 cluster members for MACS1115+01, of which 14 are emission line galaxies.  The radial velocity distribution of selected cluster members is represented by the blue bars in Figure \ref{fig:rd_dis}. The normality in velocity dispersion has been used to characterize the dynamically relaxed state of a galaxy cluster \citep[e.g.,][]{Yahil1977, Ribeiro2013}. A one sided Kolmogorov--Smirnov (KS) test yields a probability of $55\%$ for the velocity distribution of MACS1115+01 to be drawn from the normal distribution, indicating the dynamically relaxed state of the galaxy clusters.  The  velocity dispersion for MACS1115+01 is $\sigma_{cl} = 960 \pm147\ {\rm\ km/s}$. Assuming virialization, the total mass of MACS1115+01 is $ M_{\rm vel} = 1.91 \pm 0.61 \times 10^{15}\  M_{\odot}/h$, which is within the $1\sigma$ error of the  virial mass calculated from the gravitational lensing observations \citep[$ M_{\rm vir} = 1.13\times 10^{15}\  M_{\odot}/h$,][Table \ref{tb:clus_info}]{Merten2014}.  We find no obvious dependence of the stellar mass of  cluster member galaxies on the position around the cluster centre, for both emission line (Figure \ref{fig:sp_dis}: left panel - diamonds) and absorption line (Figure \ref{fig:sp_dis}: left panel - circles) systems.

For RXJ1532+30, we identify 36 cluster members, from which 21 are emission line galaxies. The cluster member mass of RXJ1532+30 also appears to be independent of the spatial position around the central BCG (Figure \ref{fig:sp_dis}: right panel).  For RXJ1532+30, the probability of the radial velocity distribution (Figure \ref{fig:rd_dis}: red bars) to be drawn from a normal distribution is only 5\% according to the KS-test. The observed velocity dispersion for RXJ1532+30 is $\sigma_{cl} = 1487  \pm 213\ {\rm\ km/s}$. Assuming a virialized state for the cluster RXJ1532+30, the total cluster mass estimated  from the velocity dispersion would be $ M_{\rm vel} = 3.78 \pm 0.73 \times 10^{15}\ M_{\odot}/h$, nearly 6 times the mass calculated from the gravitational lensing  observations \citep[$M_{\rm vir} = 0.64\times 10^{15}\  M_{\odot}/h$,][Table \ref{tb:clus_info}]{Merten2014}.  We discuss the discrepancy in the  mass estimation from the radial velocity dispersion and gravitational lensing  in Section \ref{sec:dis_grad}.

\subsection{Mass-metallicity relation (MZR)} \label{sec:mzr}

\begin{figure*}
\centering
\tiny
\includegraphics[scale=0.5, trim=0.0cm 0.0cm 0.0cm 0.0cm,clip=true]{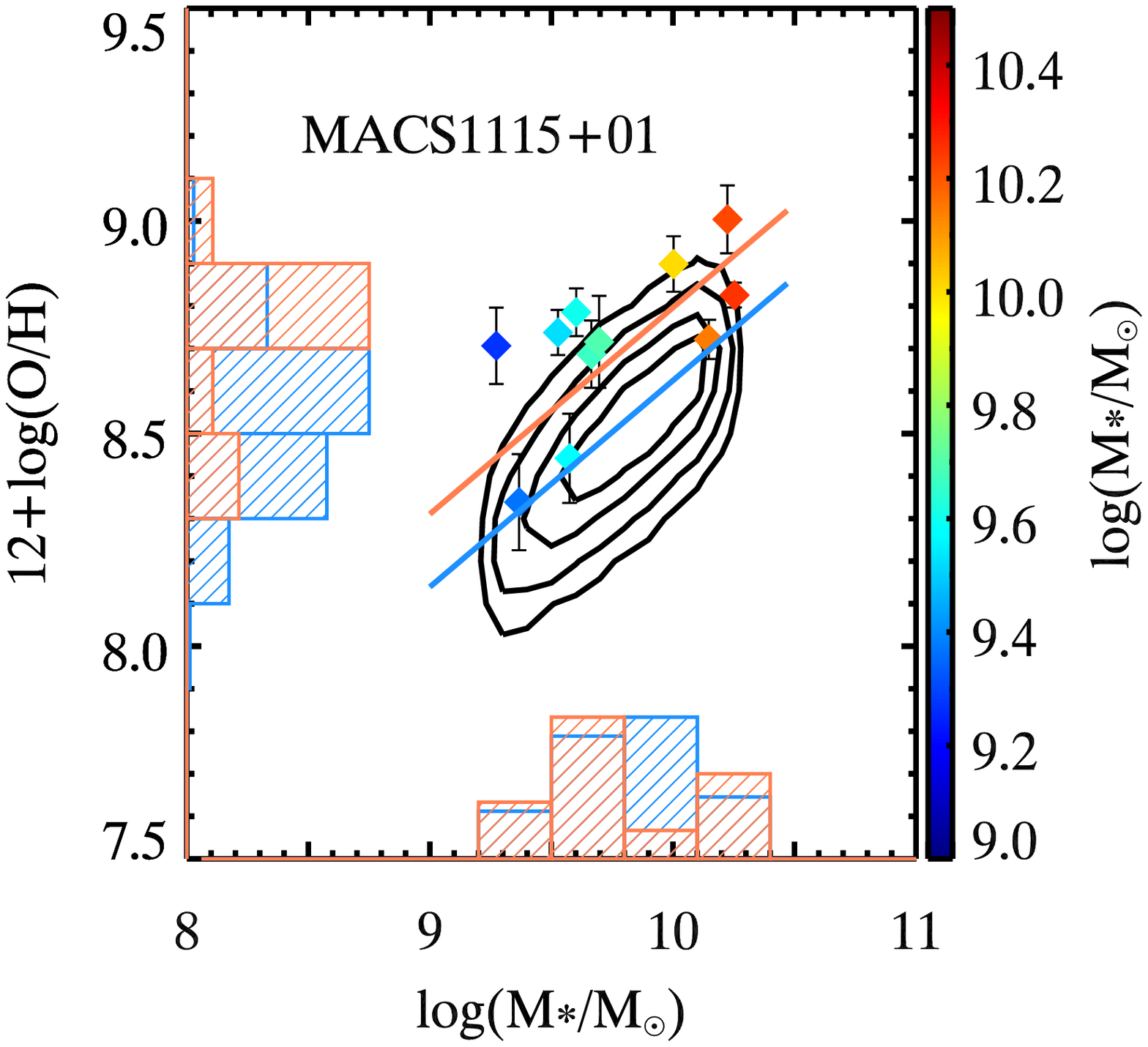}
\includegraphics[scale=0.5, trim=0.0cm 0.0cm 0.0cm 0.0cm,clip=true]{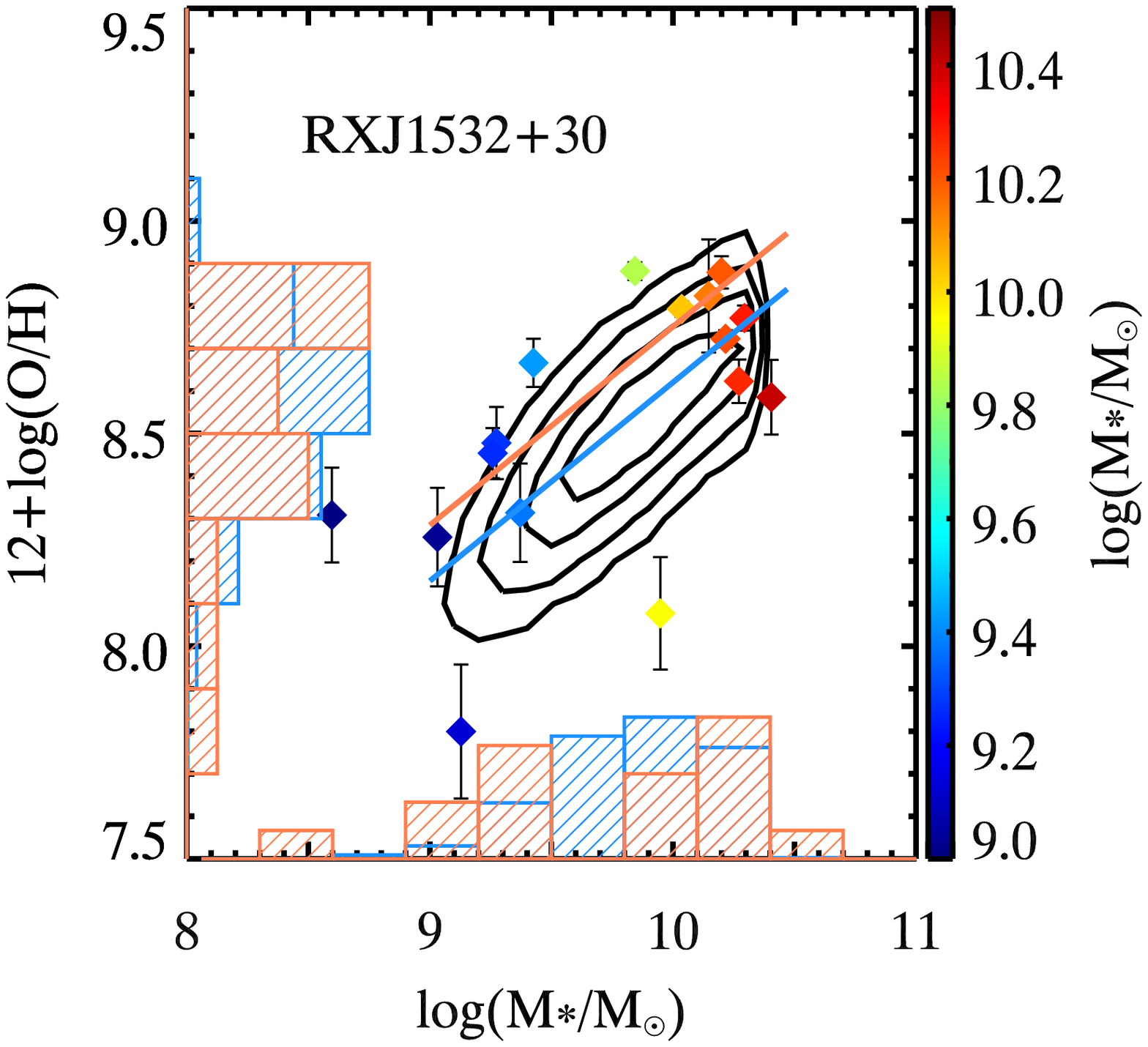}
\caption{ The mass-metallicity relation for  cluster members. {\bf Left panel:} MACS1115+01, and {\bf right panel:} RXJ1532+30. The MZR for the local comparison sample is represented as the contours (25\%, 50\%, 70\% and 90\%) in each panel.  The red and blue curve represents the best linear fit to the  MZR to cluster and SDSS data respectively. The red and blue histogram  in each panel represents the mass and metallicity distribution of the cluster members and SDSS control sample respectively.  The metallicity distribution for MACSJ1115+01 and the local comparison sample
is clearly different.}
\label{fig:mz_dis}
\end{figure*}

In Figure \ref{fig:mz_dis}, we show the MZR for the star forming cluster members in the two clusters (colored diamonds) and we compare the cluster MZR with the MZR for the local sample (contours). The metallicities of cluster members and SDSS sample were derived using the strong emission line diagnostics $\mathrm{log([NII]/H\alpha)}$ and $\mathrm{log([NII]/[SII])}$  (See Section \ref{sec:met}). The stellar mass derived from SDSS photometry (Section \ref{sec:mass}).  After applying a S/N cut of 3 for the necessary emission lines,  the  metallicities could be calculated for 12 and 16 cluster members for  MACS1115+01 and RXJ1532+30 respectively. 

Both clusters show a positive correlation between stellar mass and gas phase metallicity. The MZR for  MACS1115+01 (Figure \ref{fig:mz_dis}: left panel) has an offset of 0.2 dex to higher metallicity compared to the local comparison sample  at a fixed mass.  A double two-sided KS-test yields the probability of stellar mass and metallicity distribution of the cluster members and the local comparison sample to be derived from the same parent population as $28.9$\% and $0.03$\% respectively for MACS1115+01.  

The MZR for RXJ1532+30 (Figure \ref{fig:mz_dis}: right panel) follows nearly the same distribution as the SDSS local comparison sample. The median offset in metallicity for RXJ1532+30 and the SDSS sample is within the errors of our observations. The KS-test probability of stellar mass and metallicity distribution of cluster members and the local comparison sample to be derived from the same parent population is $7.6$\% and $28.9$\% respectively. This implies that the mass-metallicity distribution of RXJ1532+30 and the local comparison sample are subsamples  of the same parent population.

 We perform a linear fit to derive the MZR relation for our data because the turnover mass in \cite{Zahid2014} model is poorly constrained for our data due to the stellar mass cut at ${\rm \log(M_*/M_{\odot}) < 10.52 }$. The MZR for clusters (Figure \ref{fig:mz_dis}: red curve) was fit using the same slope as derived from the SDSS sample (Figure \ref{fig:mz_dis}: blue curve).  The intrinsic scatter in the MZR of the SDSS sample is 0.13 dex. We observe an offset of 0.2 dex between the mean offset in metallicity of MACS1115+01 and the local comparison sample (Figure \ref{fig:mz_dis}: left panel). Student's t-test shows the probability of an offset of 0.2 dex by chance in the metallicity of MACS1115+01 is less than 1\%. In contrast, for RXJ1532+01 the mean metallicity offset is only 0.03 dex and the Student's t-test probability of a chance offset is 88\%, confirming the statistical insignificance of the offset.  Thus, the  cluster members for MACS1115+01 show enhanced metallicity at a fixed stellar mass, implying the cluster environment  plays a significant role in driving the chemical evolution of cluster galaxies.

The metallicity offset  in MACS1115+01 ($0.2$ dex) is significantly higher than the metallicity offset observed in any previous studies of the MZR in the field versus cluster galaxy sample. In contrast, the MZR for RXJ1532+30 is consistent to the local comparison sample. Earlier studies of the MZR in cluster versus field galaxy samples constrain the environmental dependence to be around $0.04$ dex \citep{Ellison2009}, significantly less than the metallicity offset observed for MACS1115+01. However, the cluster galaxy sample for  studying the environmental dependence of the MZR is usually collected from a large number of clusters without considering the internal variations among the galaxy clusters. Combining the metallicity data from multiple clusters  might be responsible for the non-detection of  a metallicity difference between field and cluster galaxies in previous studies. The extremely different MZR for MACS1115+01 and RXJ1532+30 in our observations supports this hypothesis.

\subsection{The metallicity gradient in galaxy clusters}

\begin{figure*}
\centering
\tiny
\includegraphics[scale=0.5, trim=0.0cm 0.0cm 0.0cm 0.0cm,clip=true]{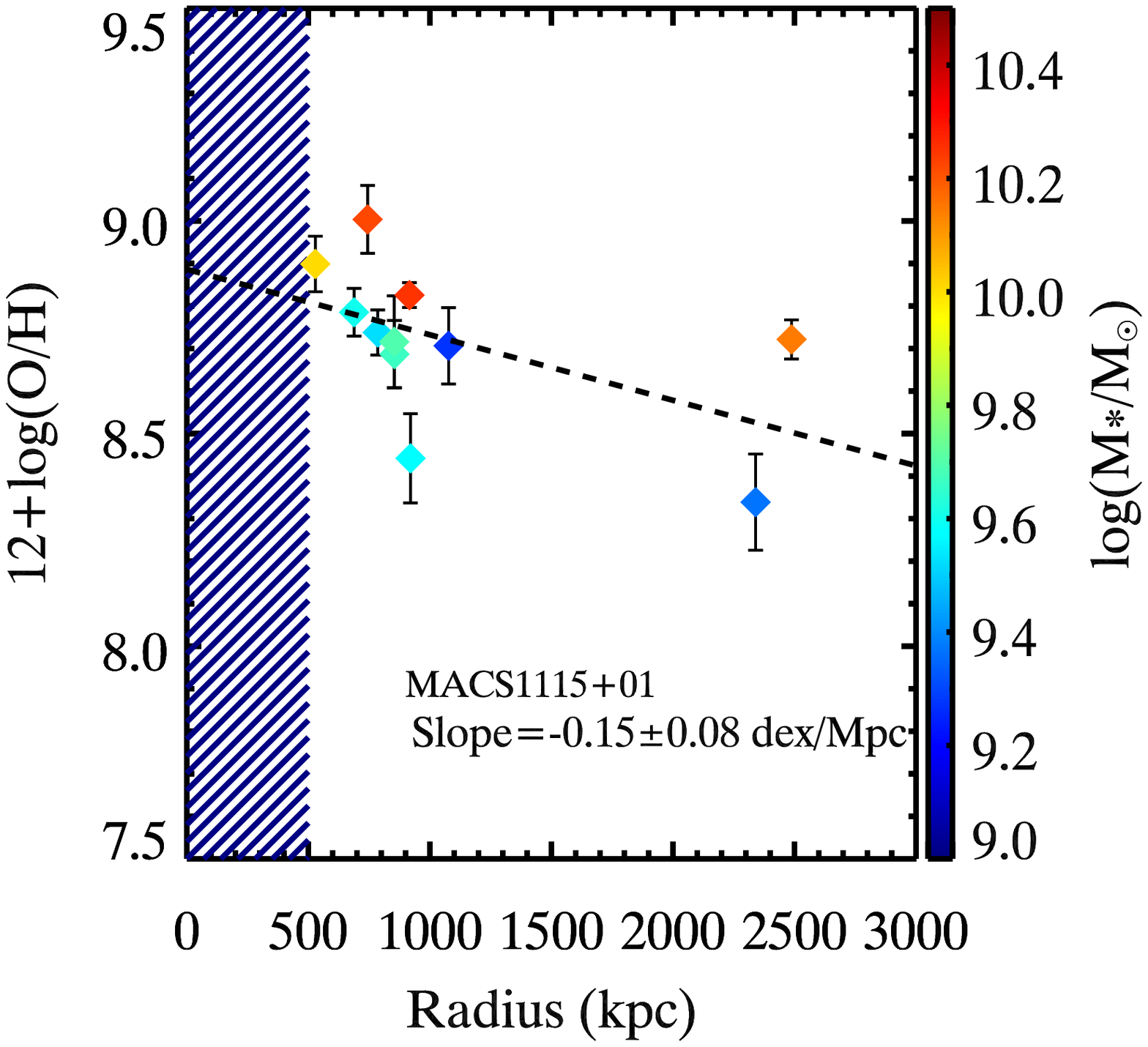}
\includegraphics[scale=0.5, trim=0.0cm 0.0cm 0.0cm 0.0cm,clip=true]{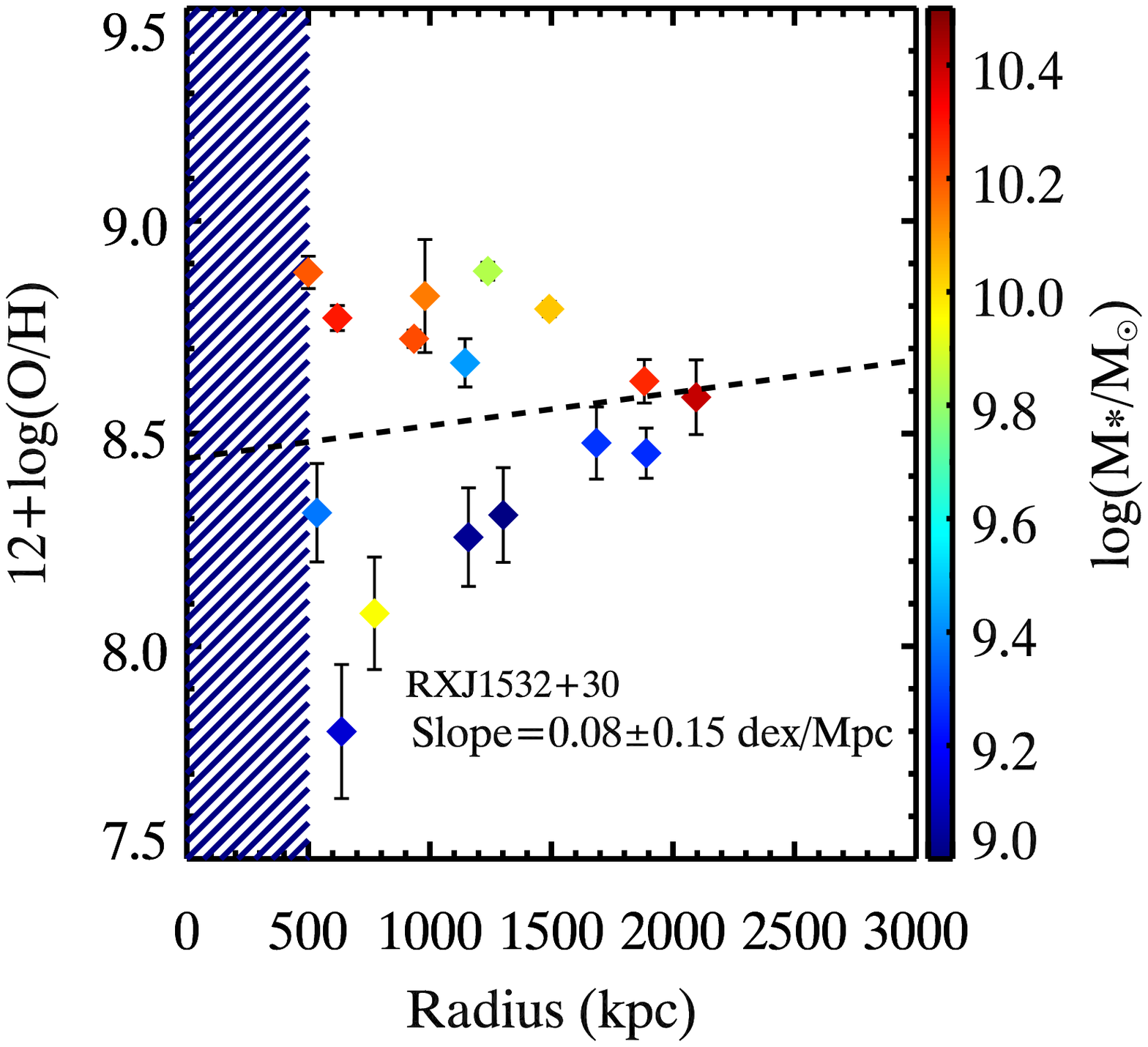}
\caption{Radial metallicity distribution  for individual clusters, where metallicity is derived using N2S2 method described in Section \ref{sec:met}.  {\bf Left panel:} MACS1115+01, and {\bf right panel:}  RXJ1532+30.  The dashed line in each figure is the best linear fit to the data. The slope and standard error in slope of the best-fitting line are labelled in each figure. The metallicity of cluster members in MACS1115+01  anti-correlates with the projected distance, whereas the radial metallicity distribution of RXJ1532+30 is bimodal that is collectively equivalent to a flat metallicity gradient.  Our GLOW-AO survey is biased against the observation of cluster member galaxies in the central part of the cluster, therefore we have shaded the central 500 kpc of the cluster (Section \ref{sec:mask}).   } 
\label{fig:z_dis}
\end{figure*}

We observe an anti-correlation between the cluster member metallicity and projected distance for MACS1115+01 (Figure \ref{fig:z_dis}: left panel).    We derive a metallicity gradient derived of $-0.15 \pm 0.08\ \mathrm{dex/Mpc}$ for MACS1115+01, after a resistant linear regression. The Spearman rank correlation coefficient between metallicity and the projected distance for MACS1115+01 is $\rho = -0.682$ with a significance of  $0.021$. Therefore, the Spearman rank probability of no correlation between the metallicity and the projected distance is 2.1\% for MACS1115+01.  The observed metallicity gradient is unlikely to be  driven by the stellar mass of galaxies because we do not observe correlation between stellar mass and projected distance (Spearman's  rank  of  $-0.25$, corresponding to a significance of 0.8$\sigma$).  To further disentangle the effect of mass and environment on the metallicity gradient, we analyze the offset in metallicity from the mass-metallicity relation. The metallicity offset also shows a negative correlation with projected distance for MACS1115+01, albeit within the limitation of our small sample size (See Appendix \ref{sec:met_offset}).

 To quantify the probability of observing a negative abundance gradient due to the intrinsic scatter in the MZR (0.13 dex), we bootstrap the data to generate a sample of cluster galaxies with a random stellar mass distribution in the cluster. We assume  a Gaussian distribution with a sigma of 0.13 dex for the offset from the mass-metallicity relation. The probability of observing a metallicity gradient of $-0.15$ dex/Mpc caused by  the 0.13 dex scatter in the MZR is only 0.1\%. Therefore, the observed metallicity gradient is unlikely to be driven by the intrinsic scatter in the MZR. 

The metallicity versus projected distance distribution  for RXJ1532+30 (Figure \ref{fig:z_dis}: right panel) is different from MACS1115+01.  A slight positive gradient ($0.08 \pm 0.15\ \mathrm{dex/Mpc}$) was derived  for RXJ1532+30  using robust linear regression. The Spearman rank correlation coefficient for RXJ1532+30 is $\rho = -0.044$ with a significance of 0.87 i.e., the probability of no correlation between the metallicity and the projected distance is $87$\%.  Thus, the metallicity gradient in RXJ1532+30 is statistically flat considering all observed cluster members. 

\begin{figure}
\centering
\tiny
\includegraphics[scale=0.5, trim=0.0cm 0.0cm 0.0cm 0.0cm,clip=true]{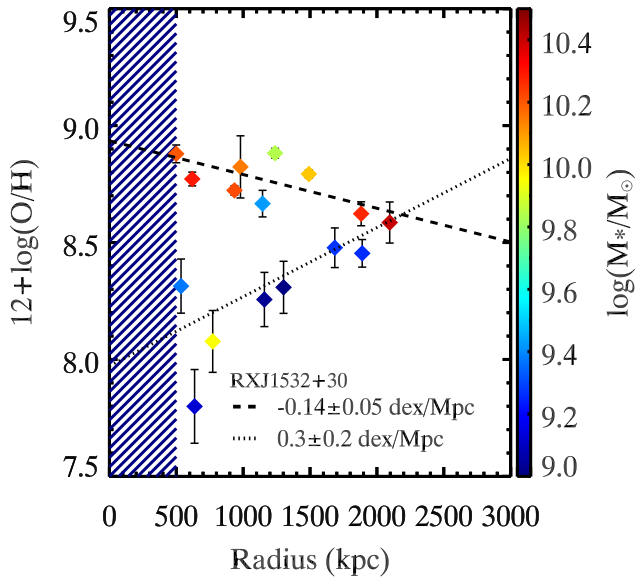}
\caption{Bimodal metallicity distribution of RXJ1532+30 using a stellar mass cut of $ {\rm \log(M_*/M_{\odot}) = 9.5}$. The dashed line represents the best-fitting  metallicity gradient  line for galaxies with ${\rm \log(M_*/M_{\odot}) > 9.5}$ (upper branch) and the dotted line gives the best-fitting metallicity gradient line for galaxies with ${\rm  \log(M_*/M_{\odot}) < 9.5}$ (lower branch).  The figure shows a negative abundance gradient  for high mass galaxies with slope similar to MACS1115+01 (Figure \ref{fig:z_dis}: left panel), whereas low mass galaxies show a positive abundance gradient.  }
\label{fig:bimodal}
\end{figure}

Interestingly, the metallicity versus projected distance for RXJ1532+30 seems to show two branches.   If we exclude the low-mass galaxies ($\log(M_*/M_{\odot}) < 9.5$), the linear regression for the rest of the cluster members yields a negative metallicity gradient of $-0.14 \pm 0.05\ {\rm dex/Mpc}$ (Figure \ref{fig:bimodal}: dashed line; hereafter upper branch). Similar linear regression fitting for the low-mass galaxies ($\log(M_*/M_{\odot}) > 9.5$) yields a positive abundance gradient  of $0.3 \pm 0.2\ {\rm dex/Mpc}$ (Figure \ref{fig:bimodal}:  dotted line; hereafter lower branch). Our observations indicate  that the lower-mass galaxies constituting the positive radial gradient are either interlopers or a group of  in-plane mergers from a sub-cluster (further discussed in Section \ref{sec:dis_grad}).

\section{Discussion }\label{sec:discussion}

We present the first detections of negative radial gradients in the gas-phase metallicity of galaxies in cluster environment.  We survey two CLASH clusters, MACS1115+01 and
RXJ1532+30, with the DEIMOS/Keck-II detecting 22 and 36 cluster members for each respectively.  The gas-phase metallicities are derived for a sample of 12 and 16 star-forming galaxies in MACS1115+01 and RXJ1532+30 respectively, using strong emission line diagnostics.

\subsection{Gas-phase metallicity gradient and cluster dynamics}\label{sec:dis_grad}

Even though MACS1115+01 and RXJ1532+30 are both massive clusters with $ M_{\rm vir} \sim 10^{14}\ M_{\odot}/h$, the observed metallicity distribution is significantly different. The gas-phase metallicities of star-forming galaxies in MACS1115+01 exhibit an anti-correlation with projected distance, with a gradient of $-0.15 \pm 0.08\ \mathrm{dex/Mpc}$  (Figure \ref{fig:z_dis}: left panel), whereas the metallicity distribution of RXJ1532+30 with projected distance  is suggested to be bimodal (Figure \ref{fig:z_dis}: right panel).    In addition, the MZR of MACS1115+01 exhibits an elevated metallicity by 0.2 dex  for cluster members as compared to the local SDSS sample (Figure \ref{fig:mz_dis}: left panel). In contrast, the MZR of RXJ1532+30  is statistically consistent with the local comparison sample (Figure \ref{fig:mz_dis}: right panel). 

Contrary to X-ray morphology studies that conclude the relaxed nature of both clusters \citep{Allen2007}, the redshift distribution from our study indicates that the two clusters have different dynamical states.  A standard KS-test suggests the redshift distribution of RXJ1532+30 is non-Gaussian as opposed to the Gaussian redshift distribution of MACS1115+01. The comparison of virial mass estimated from the velocity dispersion measurement of our observations to the virial mass estimated from  weak lensing  also suggests differences in the dynamical state.    From the velocity dispersion, we calculate the virial mass for MACSJ1115 to be $M_{\rm vel} = 1.91 \pm 0.61 \times 10^{15}\  M_{\odot}/h$,  consistent with the lensing mass within the 1-sigma errors \citep[$ M_{\rm vir} = 1.13 \pm 0.1 \times 10^{15}\  M_{\odot}/h$,][]{Merten2014}.    For RXJ1532, the estimated virial mass is $ M_{\rm vel} = 3.78 \pm 0.73 \times 10^{15}\ M_{\odot}/h$ from velocity dispersions, about 6 times larger than the lensing mass \citep[$ M_{\rm vir} = 0.64 \pm 0.09 \times 10^{15}\  M_{\odot}/h$,][]{Merten2014}.    In fact, the presence of radio mini-halos and the inhomogeneity in large scale X-ray temperature distributions in RXJ1532 also indicate  possible unrelaxed dynamics \citep{Kale2013, Hlavacek-Larrondo2013}.

The two branches in  the radial metallicity distribution of RXJ1532+30 could be the result of a minor merger with a cooler sub-cluster or a group of interloper galaxies.  The upper branch ($ \log(M_*/M_{\odot}) > 9.5$) cluster members of  RXJ1532+30  exhibit a negative abundance gradient with a slope  matching the metallicity gradient of MACS1115+01 (Figure \ref{fig:z_dis}, Figure \ref{fig:bimodal}). However, the lower branch galaxies ($ \log(M_*/M_{\odot}) < 9.5$) exhibit a positive metallicity gradient. 

The existence of  low mass galaxies in a separate substructure is confirmed by a virial mass analysis. We measure a velocity dispersion of  $773.6 \pm 162.3$  km s$^{-1}$ for RXJ1532+30 after excluding the lower branch galaxies, which estimates the  virial mass for RXJ1532+30 to be $ M_{\rm vel} = 1.02 \pm 0.55 \times 10^{15}\  M_{\odot}/h$. The estimated virial mass after excluding the lower branch galaxies  is consistent with the lensing mass within 1-sigma errors, suggesting that the low-mass galaxies belong to a separate substructure and/or are a population of interloper galaxies.  The possibility of a minor merger between a cooler sub-cluster and RXJ1532+30 was first suggested by \cite{Hlavacek-Larrondo2013} based on the spatial asymmetry in the large scale X-ray temperature of RXJ1532+30. We hypothesize that the upper branch galaxies ($ \log(M_*/M_{\odot}) > 9.5$) belong to the dynamically relaxed massive component of the cluster RXJ1532+30, with a metallicity gradient similar to the dynamically relaxed cluster MACS1115+01.  The lower branch galaxies belonging to a merging cooler sub-cluster and/or interloper galaxies that are misidentified as cluster galaxies due to projection effects. To understand  whether projection effects play a role in our metallicity gradient measurements, we compare our results with the metallicity gradient  from  Rhapsody-G simulations.

\subsection{Comparison with Rhapsody-G simulations}

We compare our data to the {\sc Rhapsody-G} simulations performed with the {\sc Ramses} \citep{Teyssier2002} adaptive mesh refinement (AMR) code (\citealp{Hahn2015}, \citealp{Martizzi2015}, \citealp{Wu2015}) because the mass rage of simulated clusters matches  the mass of observed clusters. The {\sc Rhapsody-G} sample currently includes cosmological hydrodynamical zoom-in simulations of 10 massive galaxy clusters of virial mass $M_{\rm vir}\sim 6\times 10^{14}$ M$_{\odot}/h$. The simulations assume standard $\Lambda {\rm CDM\ cosmology\ with}\ \Omega_{\rm M}=0.25$, $\Omega_{\rm \Lambda}=0.75$, $\Omega_{\rm b}=0.045$, $h=0.7$. The different cosmological parameters of the observations and simulations will introduce an offset in distance measurement up to 20 kpc.  In the {\sc Rhapsody-G} terminology, the simulations used for our comparison have `R4K resolution': dark matter particle mass $m_{\rm dm}=8.22 \times 10^{8} h^{-1}$ M$_{\odot}$; initial baryonic matter resolution element $m_{\rm b}= m_{\rm dm} \Omega_{\rm b}/(\Omega_{\rm M} - \Omega_{\rm b})=1.80 \times 10^{8} h^{-1}$ M$_{\odot}$; maximum resolution reached in these simulations is $\Delta x = 3.8 h^{-1}$ kpc (in physical units). Gas cooling, star formation, metal enrichment from supernovae, supernova feedback and AGN feedback are included in {\sc Rhapsody-G}. 

These simulations successfully reproduce the mass distribution, baryonic abundances, cool core/non-cool core dichotomy and SZ properties of observed clusters \citep{Hahn2015}.   We select simulated cluster member galaxies with star formation rate  greater than $1 M_{\odot}\ {\rm yr^{-1}}$ to best compare with the observations of metallicity gradient. Note that the absolute metallicity of the simulations is lower than that of observations, therefore we can not  directly compare the absolute simulated and observed metallicities. This offset may be related to the observational metallicity calibration used \citep[e.g.,][]{Kewley2008}. We do not compare the impact of environment on the mass-metallicity relation in simulations because there are no field galaxies in the simulations. However, we can investigate signatures of relative metallicities, such as metallicity gradients, in the Rhapsody-G simulations. We note that a direct comparison with the simulations of \cite{Taylor2015}, who use a sophisticated supernova feedback and chemical evolution prescription, was not possible because their current simulations do not contain relaxed clusters on the mass scales of interest here.

\begin{figure}
  \centering
  \tiny
\includegraphics[scale=0.5, trim=0.0cm 0.0cm 0.0cm 0.0cm,clip=true]{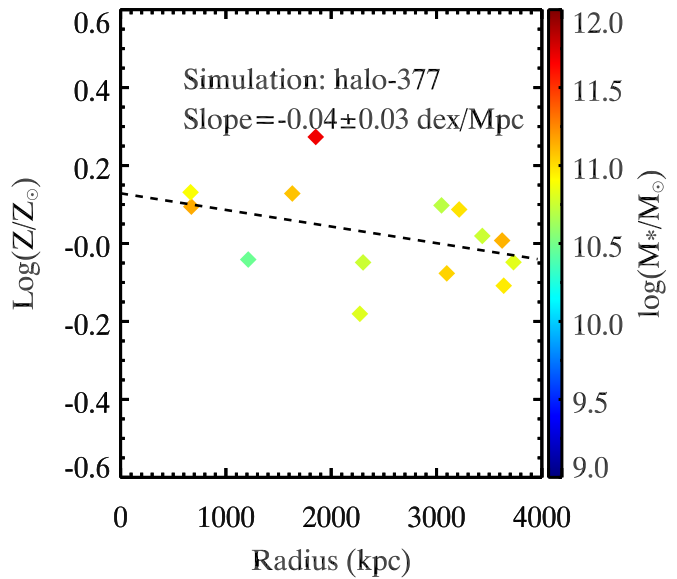}
\caption{Radial distribution of star-forming galaxies in halo-377 in the Rhapsody-G simulations. We have used projected distance instead of radial distance and colour coded the simulated galaxies according to their stellar mass for a direct comparison with the observations.  The simulations show mass-independent negative metallicity gradient similar to the metallicity gradient in our  observations.}
\label{fig:sim_clus}
\end{figure}

We select the simulated galaxy cluster, which has the closest value of concentration to the observations.  The concentration is a proxy for the age and dynamical state of the cluster \citep{Dutton2014}.  Therefore, we use the concentration parameter of the simulations ($c_{\rm vir} = r_{\rm vir}/r_s$, where the $r_{\rm vir}$ is the virial radial and $r_s$ is the scale radius of the density profile of the cluster)  to match with the  observations. Both MACS1115+01 and RXJ1532+30 have similar concentration parameters resulting in the selection of the halo-377 for both clusters. Observationally, the similar concentration of MACS1115+01 and RXJ1532+30 does not fully rule out the dynamically unrelaxed nature of RXJ1532+30 because concentration is not strongly affected by  minor merger events.  

The comparison simulated cluster exhibits a negative abundance gradient  of $-0.04 \pm 0.03\ {\rm dex/Mpc}$ (Figure \ref{fig:sim_clus}). The negative metallicity gradient in the simulation is not driven by the stellar mass  in accordance  with the mass-independent abundance gradient of our  observations.  The existence of a negative metallicity gradient in simulations confirms that the observed metallicity gradient is not a projection effect, but the shallower slope may be related to a lack  of information about the projection axis of the observed clusters.  Our future work will include a comprehensive analysis of the projection effects on the observed metallicity gradient using simulations (Gupta et al., in prep.). 

\begin{figure*}
  \centering
  \tiny
 \includegraphics[scale=0.5, trim=0.0cm 0.0cm 0.0cm 0.0cm,clip=true]{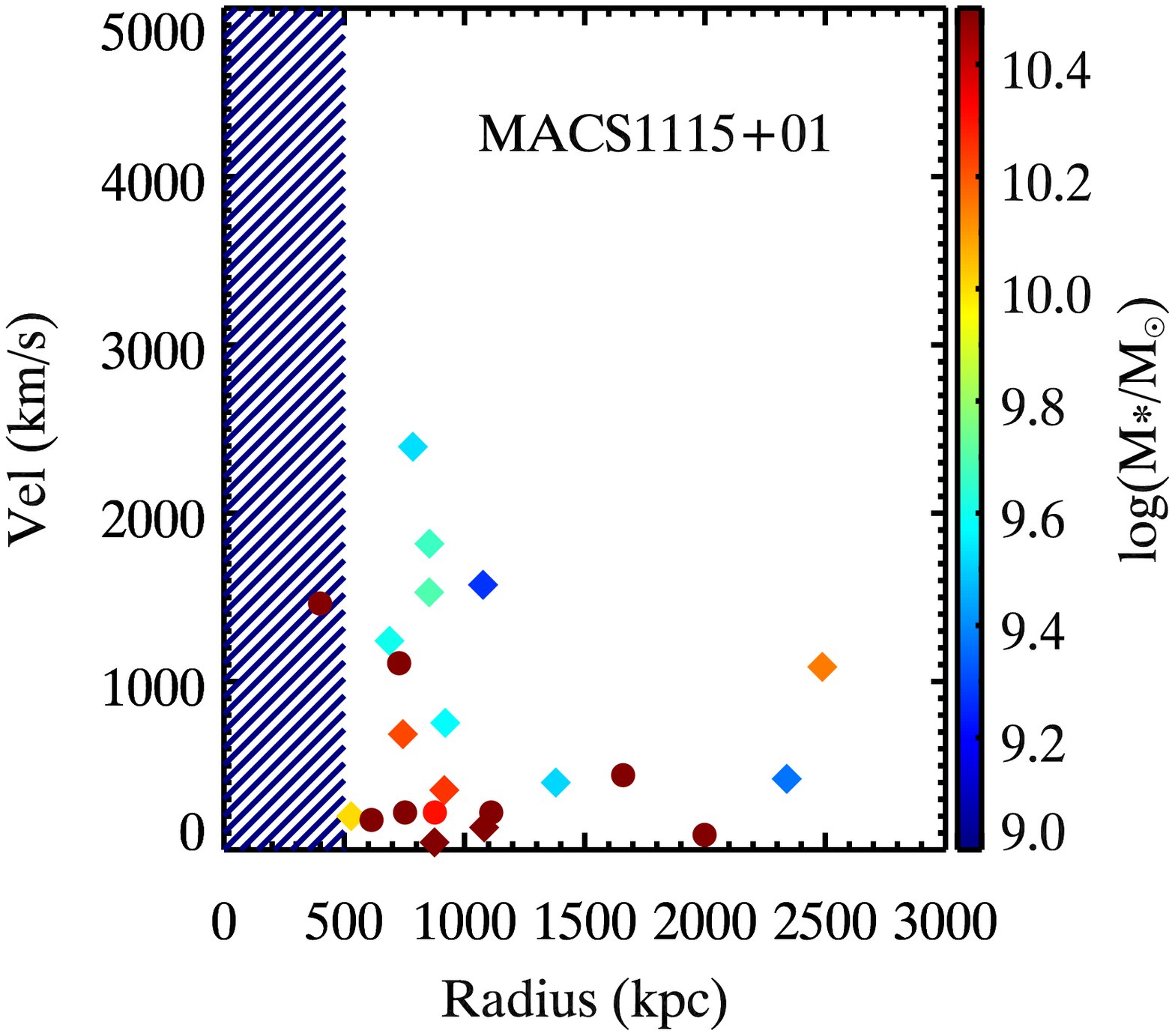}
\includegraphics[scale=0.5, trim=0.0cm 0.0cm 0.0cm 0.0cm,clip=true]{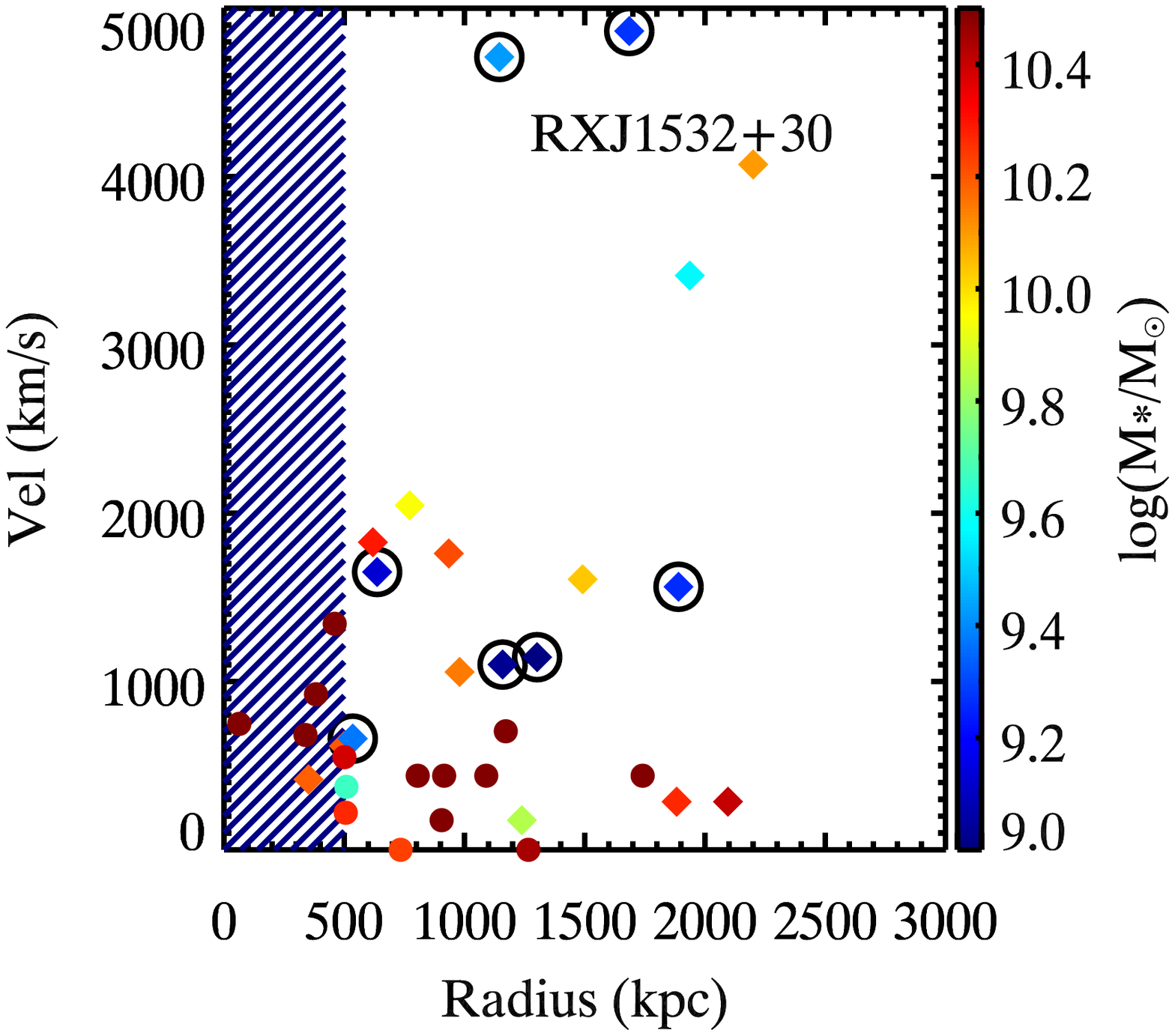}
\caption{Projected phase space diagram for the observed clusters. {\bf Left panel:} MACS1115+01, and {\bf right panel:} RXJ1532+30. Emission line galaxies are represented as diamonds and absorption line galaxies as circles. The data is color coded with stellar mass. There is inconclusive evidence of mass segregation from the projected phase space diagram for the two clusters. The open circles in the right panel represent galaxies that belong to the lower branch (${\rm \log(M_*/M_{\odot}) < 9.5 }$) in Figure \ref{fig:bimodal}. The random positions of the merging sub-cluster galaxies suggests an in-plane merger.}
\label{fig:phase_space}
\end{figure*}

\subsection{Origin of cluster-scale metallicity gradient}

\subsubsection{Truncation of galactic-disk}

 A galaxy moving through the dense environment of the galaxy cluster experiences galaxy harassment and ram pressure stripping, resulting in truncation of the outer galactic-disk.  Although, the metallicity measurements via strong emission lines are biased towards the high surface brightness regions, the truncation of the outer-galactic disk can enhance the  bias in metallicity measurements to the inner metal-rich galactic-disk \citep[e.g.,][]{Yuan2013, Genel2016}. Therefore, as the galaxy falls into the galaxy cluster, a consecutive decrease in the galactic-disk size as a result of the loss of outer-disk due to ram pressure stripping or galaxy harassment, would result in successive high metallicity measurements in spatially non-resolved observations.  This effect might play a role in causing the observed negative abundance gradient in the global metallicity of star-forming cluster members. 

Morphological changes in the galaxy as a function of radial distance are a direct consequence of the galactic-disk destabilization/truncation due to galaxy harassment and/or ram pressure stripping. Unfortunately, the existing imaging data from CFHT and SDSS  is not deep enough to observe these morphological changes, and HST data only spans the inner $\sim$300 kpc of the cluster. Thus, we require deeper imaging data in the outskirts of galaxy clusters  to observe the  morphological changes in star-forming galaxies with radial distance. Separating the effect of ram pressure and galaxy harassment is impossible with the current sample, as both galaxy clusters in our current sample are X-ray luminous. Future metallicity gradient observations with a broader sample of  X-ray luminous (ram pressure dominant) and the X-ray non-luminous (galaxy harassment dominant) clusters can help to distinguish the relative role of ram-pressure stripping and galaxy harassment. Finally, in order to understand the effect of disk truncation on metallicity gradients, deep imaging and spatially resolved spectroscopic observations are required.

\subsubsection{Self-enrichment via strangulation}

In addition to disk truncation due to ram pressure stripping or galaxy harassment, the ICM also strips the galaxy of the loosely bound hot/warm halo gas (fuel for future star formation). Depriving the galaxy of a fresh supply of gas results in a gradual decline in star formation as the galaxy consumes its existing cold gas reservoir  \citep[strangulation,][]{Larson1980}. The dilution  from the inflowing pristine gas in the strangulation mode does not exist, which results in self-enrichment of the galaxy as it goes through successive cycles of star formation. Assuming the infall time-scale for galaxies in inner region of the cluster is greater than the infall time-scale for galaxies in the outskirts,  the star-forming galaxies in the inner regions, which have undergone more cycles of star formation would be more metal rich compared to galaxies in outskirts. Thus, preferential self-enrichment of galaxies in the inner regions of the cluster as compared to the galaxies in the outskirts can also generate the observed negative abundance gradient. In order to study the self enrichment of galaxies in galaxy clusters due to strangulation, we require integral field spectroscopy of a mass-matched sample of star-forming cluster galaxies at different cluster-centric distances.

The stellar mass of a galaxy will also increase during self-enrichment via strangulation \citep{peng2015}.  Therefore, preferential  self-enrichment of galaxies in the inner regions of the cluster as compared to the galaxies in the outskirts should result in a direct correlation between the stellar mass and cluster-centric distance. The lack of an apparent stellar mass  segregation for emission line galaxies might be a result of one or more of the following: (i) selective removal of massive quenched galaxies due to our sample bias  towards star-forming galaxies, (ii) mass growth as a function of infall time depends on the details of star formation history, such as mergers and ram pressure-enhanced star formation rate \citep{Dressler1983, Gavazzi1995}, (iii) projection effects leads to uncertainty in radial distance measurement (discussed in the subsequent subsection).

\subsubsection{Systematic uncertainties due to projection effects}\label{sec:dis_pro}

Observational measurements of the absolute cluster-centric distance are impossible, we only have information about the projected distance of a galaxy in a triaxial cluster halo. Adding the line of sight velocity of the galaxy relative to the galaxy cluster  can reduce the uncertainty in the absolute cluster-centric distance measurement. A recent study by \cite{Maier2016} uses a projected phase space diagram to separate the cluster members into groups of accreted versus infalling galaxies. Adding the line of sight velocity via the projected phase-space diagram does not improve the signature of mass segregation for star-forming galaxies in our observations (Figure \ref{fig:phase_space}). In addition, the low mass cluster members for RXJ1532+30 (lower branch on Figure \ref{fig:bimodal})   do not occupy any special location on the projected phase-space diagram (Figure \ref{fig:phase_space}: right panel), which we propose belongs to an in-plane merging cooler sub-cluster and/or interloper galaxies. However, even the projected phase-space diagram does not contain full 3 dimensional information about a galaxy cluster, and therefore, is insufficient to rule out the uncertainty in radial distance measurements. 

Our current simulations do not have sufficient resolution to fully investigate the effects of disk truncation and self-enrichment on the origin of metallicity gradient.   High-resolution simulations would allow us to understand the change in galaxy's properties such as metallicity, stellar mass,  fraction of cold and hot gas, as the galaxy falls into the cluster environment. The existence of resolved low stellar mass galaxies in the high-resolution simulations would provide a better comparison with the observations. A more detailed analysis of galaxy tracking in simulations with high resolution is  required  to properly study the effects influencing the metallicity distribution in cluster satellites.
Moreover, to fully understand the origin of cluster-scale metallicity gradients, we require metallicity gradient observations with a broader sample of a range of X-ray luminosity and dynamical state. 

\section{Conclusions}

Using the multi-object optical spectrograph DEIMOS on the Keck II, we observe two CLASH clusters at $z\sim0.35$, MACS1115+01 and RXJ1532+30. We analyze the gas-phase metallicity of star-forming galaxies derived from [NII]/[SII] and [NII]/H$\alpha$,  as a function of projected distance from the central BCG. Our main results are: 

\begin{itemize}

\item For MACS1115+01, we identify  22 (14 - emission line galaxies) cluster members corresponding to a velocity dispersion of $\sigma_{cl} = 960 \pm147\ {\rm\ km/s}$. For RXJ1532+30, we identify 36 (21 - emission line galaxies) cluster members corresponding to a velocity dispersion of $\sigma_{cl} = 1487  \pm 213\ {\rm\ km/s}$. The virial mass estimated from the velocity dispersion is consistent with the gravitational lensing mass for MACS1115+01, but is 6 times more than the gravitational lensing mass for RXJ1532+30.

\item We observe a significant metallicity enhancement in the star-forming galaxies in MACS1115+01 ($0.2$ dex) at a fixed stellar mass range with respect to the local comparison galaxies. This is the largest metallicity enhancement observed for  galaxy clusters to date. The mass-metallicity relation of RXJ1532+30 is consistent  with the mass-metallicity relation of  local comparison galaxies. The significant difference between mass-metallicity relation of MACS1115+01 and RXJ1532+30 show that the non-detection of metallicity enhancement in galaxy clusters in previous studies might be due to the combination of data from multiple clusters with different mass-metallicity relation. 

\item We present the first observation of a negative abundance gradient in the global metallicity of star-forming galaxies in  galaxy clusters. We observe an abundance gradient of $-0.15 \pm 0.08\ {\rm dex/Mpc}$ for MACS1115+01 and a bimodal like metallicity distribution for RXJ1532+30. The upper branch galaxies in RXJ1532+30 also show a negative abundance gradient similar to MACS1115+01 ($-0.14 \pm 0.05\ {\rm dex/Mpc}$). We speculate that the lower branch galaxies in RXJ1532+30 belong to a merging sub-cluster and/or are interloper galaxies.  The velocity dispersion  after excluding the lower branch galaxies is   $773.6 \pm 162.3$  km/s, which gives a virial mass consistent with the lensing mass, suggesting a separate substructure for the lower branch galaxies. 

\item  Star-forming galaxies in a simulated galaxy cluster also exhibit a negative abundance gradient, but with a shallower slope  ($-0.04 \pm 0.03\ {\rm dex/Mpc}$).  

\item We speculate that the origin of the negative abundance gradient in galaxy cluster could be due to (i) the truncation of outer galactic-disk due to galaxy harassment or ram pressure stripping,  and (ii) the self-enrichment of galaxy after the ICM cuts off the gas supply for star formation.    

\item We hypothesize that a cluster-scale gradient in the ISM metallicity of galaxy members exists in clusters that  (i) are X-ray luminous (i.e., have enough hot ICM), and (ii) in a relatively relaxed dynamical state (i.e., no significant recent merger).

\end{itemize}

In our follow-up work, we will use the Rhapsody-G simulations at significantly higher resolution to investigate the origin of  metallicity gradients in galaxy clusters. To fully test our hypothesis, we require deep imaging and integral field spectroscopy for galaxy clusters with different X-ray and dynamical properties.

\section*{Acknowledgements}

The author thank the referee for the useful and comprehensive referee report. This work is based on data obtained at the W. M. Keck Observatory. We are grateful to the Keck Observatory staff for assistance with our observations, especially Jim Lyke, Marc Kassis and Luca Rizzi.  The Observatory was made possible by the generous financial support of the W. M. Keck Foundation. A.G.  acknowledges Prof. Micheal Cooper for his help with spec2d pipeline, I-Ting Ho for supplying the SDSS data and Melanie Kaasinen for  her help in the editing of this paper.  We thank C. J. Ma and Harald Ebeling for sharing their CFHT color images for the cluster RXJ1532+30.  L.J.K. gratefully acknowledges the support of an ARC Future Fellowship and ARC Discovery Project DP130103925. K. Tran acknowledges support by the National Science Foundation under Grant \#1410728.  PT acknowledges support from a Discovery Project  DP150104329 from the Australian Research Council. The authors wish to recognize and acknowledge the very significant cultural role and reverence that the summit of Mauna Kea has always had within the indigenous Hawaiian community. We are most fortunate to have the opportunity to conduct observations from this mountain.

\begin{deluxetable*}{lrrrrrrr}
\tablecolumns{8}
\tablewidth{0pc}
\tablecaption{Data table - Emission line galaxies for MACS1115+01\label{tb:data_table_1115}}
\tablehead{
\colhead{ID}  &
 \colhead{$z$\tablenotemark{a}} &
  \colhead{r$_{\rm pro}$\tablenotemark{b}} &
   \colhead{${\rm \log(M_*/M_{\odot})}$\tablenotemark{c}} & 
   \colhead{${\rm \log([OIII]/H\beta)}$\tablenotemark{d}} &
\colhead{${\rm \log([NII]/H\alpha)}$\tablenotemark{d}} & 
\colhead{${\rm \log([SII]/H\alpha)}$\tablenotemark{d}} & 
\colhead{${\rm \log([NII]/[SII])}$\tablenotemark{d}}
}
\startdata 
1115020-0 & 0.353 & 528.79 & 10.0 & \nodata$\pm$\nodata & -0.39$\pm$0.05 & -0.63
$\pm$0.04 & 0.23$\pm$0.06\\
1115026-0 & 0.351 & 1081.7 & 10.6 & 0.53$\pm$0.02 & -0.17$\pm$0.01 & -0.44$\pm$
0.01 & 0.28$\pm$0.01\\
1115042-0 & 0.341 & 785.53 & 9.53 & -0.14$\pm$0.03 & -0.46$\pm$0.03 & -0.54$\pm$
0.04 & 0.088$\pm$0.05\\
1115069-0 & 0.352 & 875.36 & 10.9 & -0.30$\pm$0.08 & -0.63$\pm$0.08 & -0.12$\pm$
0.02 & -0.51$\pm$0.08\\
1115072-0 & 0.349 & 920.02 & 9.57 & \nodata$\pm$\nodata & -0.67$\pm$0.09 & -0.52
$\pm$0.04 & -0.15$\pm$0.09\\
1115077-0 & 0.354 & 2340.0 & 9.37 & -0.11$\pm$0.02 & -0.65$\pm$0.1 & -0.39$\pm$
0.02 & -0.26$\pm$0.1\\
1115088-0 & 0.350 & 1379.6 & 9.51 & \nodata$\pm$\nodata & \nodata$\pm$\nodata & 
-0.035$\pm$0.2 & \nodata$\pm$\nodata\\
1115096-0 & 0.345 & 1077.0 & 9.27 & -0.60$\pm$0.1 & -0.32$\pm$0.07 & -0.34$\pm$
0.05 & 0.022$\pm$0.08\\
1115103-0 & 0.355 & 743.73 & 10.2 & \nodata$\pm$\nodata & -0.28$\pm$0.07 & -0.59
$\pm$0.07 & 0.31$\pm$0.07\\
1115112-0 & 0.360 & 854.36 & 9.66 & \nodata$\pm$\nodata & -0.29$\pm$0.05 & -0.28
$\pm$0.06 & -0.0068$\pm$0.07\\
1115116-0 & 0.346 & 688.04 & 9.60 & -0.45$\pm$0.1 & -0.29$\pm$0.04 & -0.38$\pm$
0.04 & 0.091$\pm$0.05\\
1115117-0 & 0.350 & 916.24 & 10.3 & -0.62$\pm$0.05 & -0.47$\pm$0.02 & -0.65$\pm$
0.02 & 0.18$\pm$0.03\\
1115119-0 & 0.357 & 2487.8 & 10.1 & -0.56$\pm$0.05 & -0.47$\pm$0.04 & -0.55$\pm$
0.02 & 0.076$\pm$0.04\\
1115121-0 & 0.345 & 853.26 & 9.70 & \nodata$\pm$\nodata & -0.28$\pm$0.09 & -0.31
$\pm$0.07 & 0.021$\pm$0.1\\
\enddata
\tablenotetext{a}{z represents the best-fitting redshift to the DEIMOS spectra using the emission lines.} 
\tablenotetext{b}{r$_{\rm pro}$ is the projected cluster-centric distance of the cluster members from the central BCG.}
\tablenotetext{c}{${\rm \log(M_*/M_{\odot})}$ is the stellar mass derived from the SDSS photometry (See Section \ref{sec:mass}).}
\tablenotetext{d}{ log([OIII]/H$\beta$), log([NII]/H$\alpha$), log([SII]/H$\alpha$) and log([NII]/[SII]) are the emission line ratios derived from the DEIMOS spectra (See Section \ref{sec:em_flux}).}
 
\end{deluxetable*}

\begin{deluxetable*}{lrrrrrrr}
\tablecolumns{8}
\tablewidth{0pc}
\tablecaption{Data table - Emission line galaxies for RXJ1532+30 \label{tb:data_table_1532}}
\tablehead{
\colhead{ID}  &
 \colhead{$z$\tablenotemark{a}} &
  \colhead{r$_{\rm pro}$\tablenotemark{a}} &
   \colhead{${\rm \log(M_*/M_{\odot})}$\tablenotemark{a}} & 
   \colhead{${\rm \log([OIII]/H\beta)}$\tablenotemark{a}} &
\colhead{${\rm \log([NII]/H\alpha)}$\tablenotemark{a}} & 
\colhead{${\rm \log([SII]/H\alpha)}$\tablenotemark{a}} & 
\colhead{${\rm \log([NII]/[SII])}$\tablenotemark{a}}
}
\startdata 
1532008-1 & 0.363 & \nodata & 10.5 & \nodata$\pm$\nodata & -0.23$\pm$0.04 & 
\nodata$\pm$\nodata & \nodata$\pm$\nodata\\
1532017-0 & 0.353 & 772.08 & 9.95 & 0.44$\pm$0.05 & -0.99$\pm$0.1 & -0.55$\pm$
0.03 & -0.43$\pm$0.1\\
1532023-0 & 0.357 & 979.84 & 10.1 & \nodata$\pm$\nodata & -0.22$\pm$0.1 & -0.33
$\pm$0.08 & 0.11$\pm$0.1\\
1532028-0 & 0.341 & 1144.9 & 9.43 & -0.033$\pm$0.03 & -0.40$\pm$0.05 & -0.40
$\pm$0.03 & 0.0018$\pm$0.05\\
1532032-0 & 0.357 & 1158.7 & 9.03 & \nodata$\pm$\nodata & -0.72$\pm$0.1 & -0.40
$\pm$0.03 & -0.32$\pm$0.1\\
1532034-0 & 0.357 & 1301.9 & 8.60 & 0.47$\pm$0.03 & -0.77$\pm$0.09 & -0.51$\pm$
0.05 & -0.26$\pm$0.1\\
1532066-0 & 0.364 & 351.44 & 10.2 & \nodata$\pm$\nodata & \nodata$\pm$\nodata & 
\nodata$\pm$\nodata & \nodata$\pm$\nodata\\
1532074-0 & 0.365 & 499.56 & 10.2 & -0.50$\pm$0.07 & -0.42$\pm$0.03 & -0.63$\pm$
0.02 & 0.22$\pm$0.03\\
1532080-0 & 0.365 & 535.37 & 9.37 & 0.16$\pm$0.08 & -0.70$\pm$0.1 & -0.42$\pm$
0.05 & -0.27$\pm$0.1\\
1532086-0 & 0.354 & 619.21 & 10.3 & 0.14$\pm$0.08 & -0.32$\pm$0.02 & -0.40$\pm$
0.02 & 0.086$\pm$0.03\\
1532087-0 & 0.354 & 636.69 & 9.13 & 0.40$\pm$0.02 & -1.3$\pm$0.1 & -0.62$\pm$
0.01 & -0.64$\pm$0.1\\
1532093-0 & 0.354 & 934.69 & 10.2 & 0.15$\pm$0.01 & -0.38$\pm$0.02 & -0.43$\pm$
0.01 & 0.053$\pm$0.02\\
1532098-0 & 0.361 & 1238.5 & 9.84 & -0.43$\pm$0.03 & -0.37$\pm$0.02 & -0.58$\pm$
0.009 & 0.21$\pm$0.02\\
1532102-0 & 0.355 & 1491.3 & 10.0 & -0.76$\pm$0.04 & -0.44$\pm$0.02 & -0.57$\pm$
0.009 & 0.14$\pm$0.02\\
1532110-0 & 0.340 & 1685.1 & 9.27 & -0.12$\pm$0.1 & -0.31$\pm$0.07 & -0.10$\pm$
0.06 & -0.21$\pm$0.08\\
1532116-0 & 0.355 & 1889.9 & 9.26 & 0.17$\pm$0.09 & -0.50$\pm$0.05 & -0.32$\pm$
0.03 & -0.18$\pm$0.05\\
1532118-0 & 0.347 & 1936.8 & 9.57 & 0.50$\pm$0.04 & \nodata$\pm$\nodata & -0.50
$\pm$0.1 & \nodata$\pm$\nodata\\
1532119-0 & 0.361 & 1882.2 & 10.3 & \nodata$\pm$\nodata & -0.31$\pm$0.04 & -0.25
$\pm$0.04 & -0.065$\pm$0.05\\
1532122-0 & 0.361 & 2095.5 & 10.4 & -0.022$\pm$0.1 & -0.38$\pm$0.08 & -0.30$\pm$
0.04 & -0.083$\pm$0.08\\
1532125-0 & 0.343 & 2200.3 & 10.1 & -0.56$\pm$0.05 & \nodata$\pm$\nodata & 
\nodata$\pm$\nodata & \nodata$\pm$\nodata\\
\enddata

\tablenotetext{a}{Same as Table \ref{tb:data_table_1115}, but for galaxy cluster RXJ1532+30.}

\end{deluxetable*}%

\begin{deluxetable*}{lrrr}
\tablecolumns{4}
\tablewidth{0pc}
\tablecaption{Data table - Absorption line galaxies for MACS1115+01 \label{tb:data_table_1115_abs}}
\tablehead{
\colhead{ID}  &
 \colhead{$z$\tablenotemark{a}} &
  \colhead{r$_{\rm pro}$\tablenotemark{b}} &
   \colhead{${\rm \log(M_*/M_{\odot})}$\tablenotemark{c}} 
   }
\startdata 

1115025 & 0.347 & 728.38 & 10.7\\
1115043 & 0.354 & 1659.6 & 10.5\\
1115057 & 0.359 & 399.22 & 10.6\\
1115064 & 0.351 & 613.48 & 10.9\\
1115076 & 0.351 & 1111.1 & 10.5\\
1115098 & 0.353 & 876.71 & 10.3\\
1115100 & 0.352 & 1998.4 & 11.1\\
1115115 & 0.351 & 752.85 & 11.4\\
\enddata

\tablenotetext{a}{z represents the best-fitting redshift to the DEIMOS spectra using the absorption lines (See Section \ref{sec:abs_fit}).} 
\tablenotetext{b}{r$_{\rm pro}$ is the projected cluster-centric distance of the cluster members from the central BCG.}
\tablenotetext{c}{${\rm \log(M_*/M_{\odot})}$ is the stellar mass derived from the SDSS photometry (See Section \ref{sec:mass}).}

\end{deluxetable*}%

\begin{deluxetable*}{lrrr}
\tablecolumns{4}
\tablewidth{0pc}
\tablecaption{Data table - Absorption line galaxies for MACS1532+30 \label{tb:data_table_1532_abs}}
\tablehead{
\colhead{ID}  &
 \colhead{$z$\tablenotemark{a}} &
  \colhead{r$_{\rm pro}$\tablenotemark{a}} &
   \colhead{${\rm \log(M_*/M_{\odot})}$\tablenotemark{a}}
   }
\startdata 

1532009 & 0.359 & 63.504 & 10.7\\
1532012 & 0.359 & 338.83 & 10.5\\
1532018 & 0.361 & 904.93 & 10.7\\
1532019 & 0.364 & 915.37 & 10.8\\
1532027 & 0.360 & 1090.4 & 10.8\\
1532031 & 0.359 & 1171.7 & 10.9\\
1532065 & 0.358 & 382.43 & 10.6\\
1532069 & 0.356 & 460.26 & 10.5\\
1532071 & 0.359 & 500.72 & 10.4\\
1532075 & 0.361 & 506.15 & 10.3\\
1532077 & 0.364 & 508.77 & 9.66\\
1532088 & 0.362 & 734.37 & 10.2\\
1532090 & 0.360 & 805.38 & 10.6\\
1532100 & 0.362 & 1265.8 & 10.4\\
1532112 & 0.360 & 1740.8 & 11.1\\
\enddata

\tablenotetext{a}{Same as Table \ref{tb:data_table_1115_abs}, but for galaxy cluster RXJ1532+30.}

\end{deluxetable*}%

\bibliographystyle{aasjournal}

\begin{appendix}

\section{The offset from the mass-metallicity relation}\label{sec:met_offset}

\begin{figure*}
  \centering
  \tiny
 \includegraphics[scale=0.5, trim=0.0cm 0.0cm 0.0cm 0.0cm,clip=true]{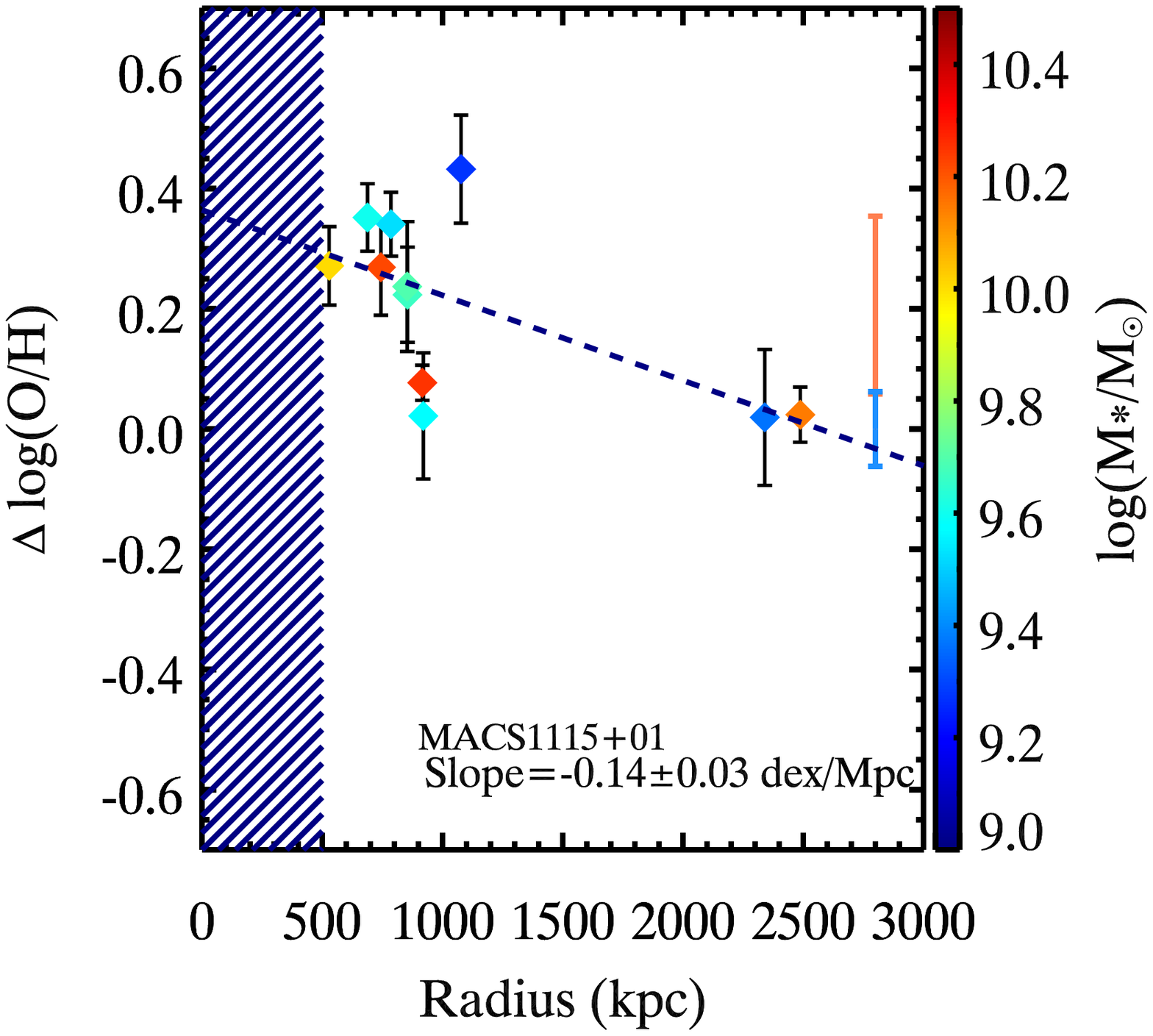}
\includegraphics[scale=0.5, trim=0.0cm 0.0cm 0.0cm 0.0cm,clip=true]{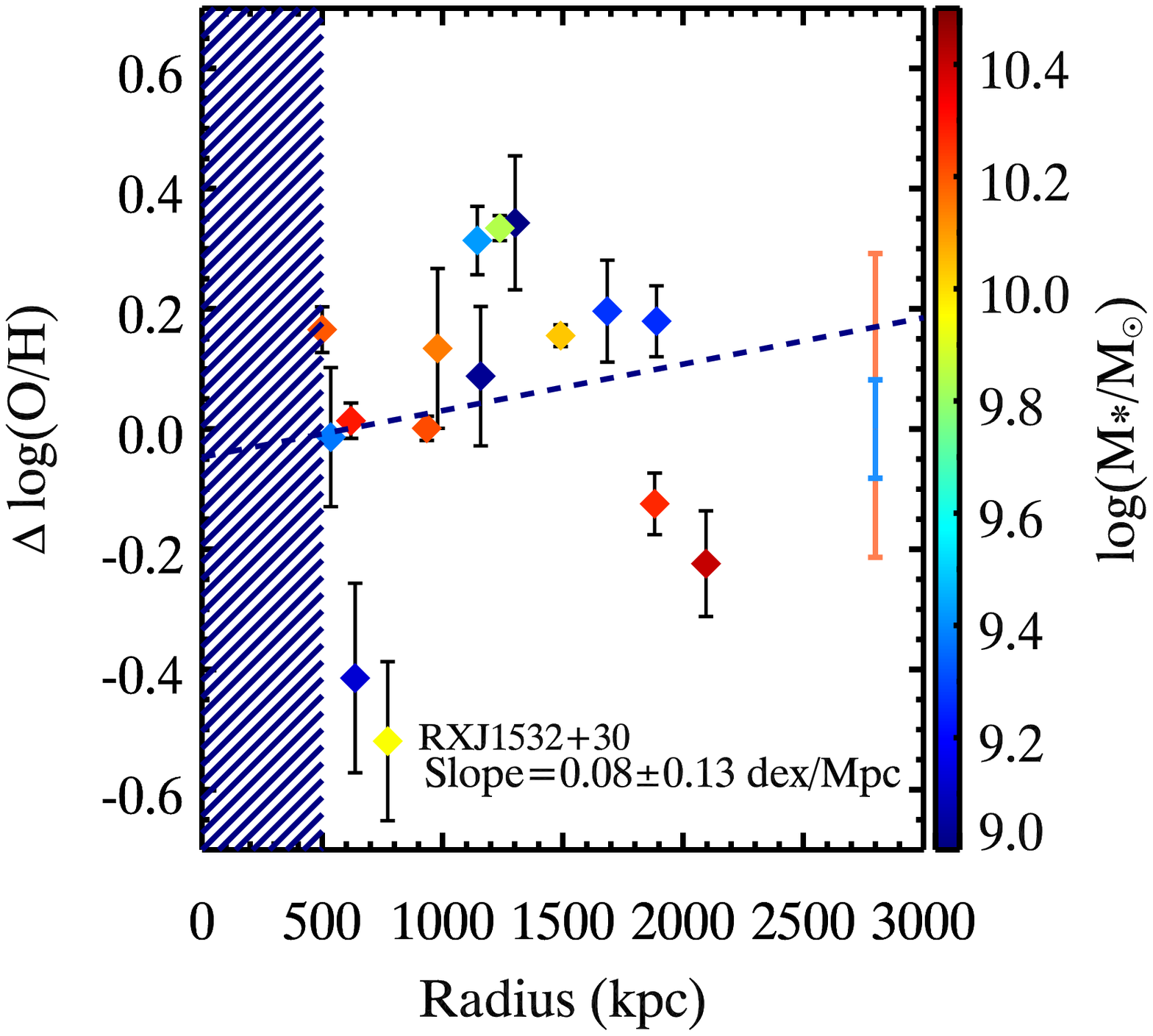}
\caption{The offset from the mass-metallicity relation of the local comparison sample (Figure \ref{fig:mz_dis}: red curve) with the projected distance. {\bf Left panel:} MACS1115+01, and {\bf right panel:} RXJ1532+30. The dashed line represents the best linear fit. The slope and standard error in slope are labelled in each figure. The blue and red error bar represents the scatter and median metallicity offset of the comparison sample and the cluster data. The offset from the MZR also exhibit anti-correlation with the projected distance for MACS1115+01, which suggests that the metallicity enhancement towards the cluster center is not driven the stellar mass. }
\label{fig:met_offset}
\end{figure*}

To separate the effect of mass and environment on the chemical evolution, we study the offset in cluster member metallicity from the mass-metallicity relation of the local comparison sample.   For MACS1115+01, cluster members in the inner parts show enhanced metallicity at a fixed stellar mass range compared to cluster members in the outskirts (Figure \ref{fig:met_offset}: left panel).  In contrast, cluster members in RXJ1532+30 do not show a correlation between the metallicity offset and the projected distance (Figure \ref{fig:met_offset}: right panel). The existence of the negative gradient in the offset from mass-metallicity relation for MACS1115+01 suggests that the cluster-scale negative abundance gradient is not driven by the stellar mass of galaxies.  However, our conclusion is limited by the small sample size of our data.

\end{appendix}

\end{document}